\documentclass[a4paper,10pt]{article}
\pdfoutput=1 
\usepackage{jcappub}
\usepackage[T1]{fontenc}
\usepackage{appendix}
\usepackage{enumitem}
\usepackage{booktabs}
\usepackage{multirow}
\usepackage{siunitx}
\usepackage{orcidlink}
\usepackage{arydshln}
\usepackage{caption}
\usepackage{subfig}
\usepackage{graphicx}

\title{\boldmath Constraining extra dimensions on cosmological scales with LISA future gravitational wave siren data}

\author[a]{Maxence Corman\orcidlink{0000-0003-2855-1149},}
\author[b]{Celia Escamilla-Rivera\orcidlink{0000-0002-8929-250X}}
\author[c]{and M.A. Hendry\orcidlink{0000-0001-8322-5405}}

\affiliation[a]{Perimeter Institute, 31 Carolina St, Waterloo, ON N2L 2Y5, Canada.}
\affiliation[b]{Instituto de Ciencias Nucleares, Universidad Nacional Aut\'onoma de M\'exico, Circuito Exterior C.U., A.P. 70-543, M\'exico D.F. 04510, M\'exico.}
\affiliation[c]{SUPA, School of Physics and Astronomy, University of Glasgow, Glasgow G12 8QQ, United Kingdom.}
\emailAdd{mcorman@perimeterinstitute.ca}
\emailAdd{celia.escamilla@nucleares.unam.mx}
\emailAdd{martin.hendry@glasgow.ac.uk}

\abstract
{The Universe is undergoing a late time acceleration. We investigate the idea that this acceleration could be the consequence of gravitational leakage into extra dimensions on cosmological scales rather than the result of a non-zero cosmological constant, and consider the ability of future gravitational-wave (GW) siren observations to probe this phenomenon and constrain the parameters of phenomenological models of this gravitational leakage.
A gravitational space interferometer such as LISA will observe massive black hole binary (MBHB) merger events at cosmological distances, and will also provide sky localization information that may permit optical and other electromagnetic (EM) surveys to identify an EM counterpart of these events. In theories that include additional non-compact spacetime dimensions, the gravitational leakage intro extra dimensions leads to a reduction in the amplitude of observed gravitational waves and thereby a systematic discrepancy between the distance inferred to such sources from GW and EM observations. We investigate the capability of LISA to probe this modified gravity on large scales, specifically in the Dvali, Gabadadze and Porrati (DGP) model. Additionally, we include a Supernova Ia sample at lower redshift in order to explore the efficacy of this cosmological probe across a range of redshifts. We then use previously published simulated catalogues of cosmologically distant MBHB merger events detectable by LISA, and which are likely to produce an observable EM counterpart.  We find that the extent to which LISA will be able to place limits on the number of spacetime dimensions and other cosmological parameters characterising modified gravity will strongly depend on the actual number and redshift distribution of sources, together with the uncertainty on the GW distance measurements. A relatively small number of sources ($\sim 1$) and high measurement uncertainties would strongly restrict the ability of LISA to place meaningful constraints on the parameters in cosmological scenarios where gravity is only five-dimensional and modified at scales larger than about $\sim 4$ times the Hubble radius. Conversely, if the number of sources observed amounts to a four-year average of $\sim 27$, then in the most favourable cosmological scenarios LISA has the potential to place meaningful constraints on the cosmological parameters -- with a precision of $\sim 1\%$ on the number of dimensions and $\sim 7.5\%$ on the scale beyond which gravity is modified, thereby probing the late expansion of the universe up to a redshift of $\sim 8$, i.e. on scales not yet tested by present EM observations.}

\begin{document}
\maketitle
\flushbottom


\section{Introduction}

The expansion of the universe
is undergoing a late time acceleration,  the reality of which was for the first time directly confirmed in the late 1990s by surveys of cosmologically distant Type Ia supernovae (SNeIa) 
and then further supported with more recent SNIa observations \cite{HST, SNLS}. Low-redshift data of baryon acoustic oscillations (BAO) \cite{BAO}, cosmic microwave background (CMB) data from the Planck Collaboration et.al \cite{Planck} and data from other cosmological probes provide further, independent support for an accelerating universe.

In the standard model of cosmology where General Relativity (GR) is the underlying framework, this acceleration may be explained by a non-zero but very small cosmological constant, $\Lambda$, that opposes the self-attraction of pressureless matter and causes the expansion of the universe to accelerate at the largest scales \cite{lambda}. Although the $\Lambda$-Cold-Dark-Matter model ($\Lambda$CDM) is very successful in explaining almost all observations, it has some theoretical issues. These include the mysterious physical origin of the two largest contributions to the energy content of the late-time universe: cold dark matter (CDM) and the cosmological constant ($\Lambda$), together with the unsatisfactory predictivity and testability of the inflation theory \cite{problem}. Therefore it is important to explore alternative explanations for the late-time acceleration of the universe.
There are several proposals in the literature in this regard, e.g. \cite{Barcenas-Enriquez:2018ili}
presented a study using data from the Hubble Space Telescope and the large-scale galaxy distribution data to demonstrate the disadvantages of $\Lambda$CDM in specific cosmic acceleration scenarios over the redshift range of these observations, thus providing the motivation to approach other modified models such as the Randall-Sundrum proposal, wherein GW data can set new constraints on extra dimensional cosmological parameters \cite{Garcia-Aspeitia:2020snv}. Therefore, 
in this work we will focus on the Dvali, Gabadadze and Porrati (DGP) braneworld model proposed in 
\cite{DGPdisc},
in which our observed four-dimensional universe is considered to be a surface (called a brane) embedded in an infinite-volume, five-dimensional space (called the bulk). This proposed model explains the late time acceleration of the expansion of the universe through a large-scale modification of gravity arising from the slow evaporation of gravitational degrees of freedom into the infinite-volume extra dimension, and without requiring a non-zero cosmological constant. The reason for first studying the DGP model is that we will borrow notions from it to then consider cosmological models including additional non-compact spacetime dimensions. 

Until now, by precisely mapping the distance-redshift relation up to a redshift of $z \sim 1$, SNeIa measurements were the most sensitive probe of the late time acceleration of the universe. Another potentially very powerful probe of the universe are GWs emitted from merging binary black holes and neutron stars. In particular, since the propagation of gravitons and photons could differ at a fundamental level, we argue here that GWs emitted by coalescing pairs of massive black holes, at the centre of distant galaxies, observed with an EM counterpart, may be used as an alternative way to test modified theories of gravity.
To measure GWs emitted by these massive black hole binary (MBHB) mergers, with high signal-to-noise ratio (SNR) and in the previously unobserved redshift range $1 < z < 8$, ESA and NASA will build the Laser Interferometer Space Antenna (LISA \footnote{http://lisamission.org/}).
Such GW sources can be thought of as \textit{standard sirens} (gravitational analogues of standard candles such as SNIe) in the sense that the GWs emitted by a compact binary directly encode information about the gravitational-wave luminosity of the binary system, and thus its luminosity distance \cite{StandardSirens}. 
Still, MBHBs have the big disadvantage that the redshift needs to be measured independently, e.g. by optically identifying the host galaxy. However, if we do successfully identify the host galaxy then one question we might be able to answer with such joint events is whether long-wavelength gravitational waves and short-wavelength EM radiation experience the same number of spacetime dimensions \cite{pardo}. In higher dimensional theories such as the DGP model, as the GWs propagate through spacetime they \textit{leak} into the extra dimensions, leading to the effect that cosmologically distant sources appear dimmer than they truly are -- hence resulting in a systematic error in the inferred distance to the gravitational wave source. Assuming, as is the case for the DGP model, that light and matter propagate in four spacetime dimensions only, EM waves remain unaffected.  Hence, if we can make an independent measurement of the luminosity distance to the GW source by measuring the redshift of its EM counterpart then this allows us to place limits on the gravitational leakage. 

In this paper we investigate the consequences of higher dimensional theories with non-compact extra dimensions, borrowing notions from the DGP model to forecast the capability of LISA to constrain theories with gravitational leakage from gravitational wave standard sirens with an observable EM counterpart. We find that LISA's ability to set bounds on the number of additional non-compact spacetime dimensions and the other parameters characterising the theory, namely the screening scale and the transition steepness, will depend strongly on the actual redshift distribution of massive black hole binary merger events, the corresponding efficiency in identifying their host galaxy and the uncertainty on these measurements. Also, to compare our forecast at high redshift, we perform a DGP test with the standard Pantheon SNeIa sample at low redshifts, in the sense that this test will be our pivot model in order to carry out a model selection dependent on the number of dimensions and screening scale.

This paper is organised as follows: in Sec.\ref{section2} we introduce the dynamics of the background metric of the universe in the DGP model. Borrowing notions from the DGP model, according to the positive branch, in Sec.\ref{sec:lumdis} we write expressions for the luminosity distance expressions for supernovae and GW standard sirens, and then describe the effects of gravitational leakage on the GW waveform in higher-dimensional theories with non-compact dimensions. 
In Sec.\ref{LISA} we investigate the predicted ability of LISA to constrain higher-dimensional gravity. First, in Sec.\ref{cat} we summarise the properties of the MBHB population catalogues that we use to simulate realistic GW data. In Sec.\ref{bayes} we then briefly review our statistical methods and perform the model selection comparison on the Pantheon SNe sample, using the DGP model as a pivot model. Finally, in Sec.\ref{resultsLISA} we present and analyse our results for the predicted LISA constraints on the parameters characterising higher-dimensional theories. In addition, we present the analyses for the joint Pantheon and LISA samples. Our main conclusions are then presented and summarised in Sec.\ref{conclusion}.


\section{Extra-dimensional theory: DGP cosmological solutions} 
\label{section2}
In the DGP braneworld theory, matter and all standard model forces and particles are pinned to a $(3+1)$-dimensional brane universe, while gravity is free to explore the full five-dimensional empty bulk. For the purposes of our analysis we are mainly interested in the geometry of the 4-D braneworld universe, which (assuming a homogeneous, isotropic expanding universe) is at all times described by a standard Friedmann-Lemaitre-Robertson-Walker (FLRW) metric 
\begin{equation}\label{FLRW}
\begin{split}
ds^2 & =- d\tau^2 +a^2(\tau)\gamma_{ij}d\lambda^id\lambda^j, \\
	& = -d\tau^2 +a^2(\tau)(dr^2 + S_k^2(r)d\psi^2), \\
\end{split}
\end{equation}
where the chosen coordinates are the cosmological time $\tau$ and the spatial comoving coordinates of the observable universe $\lambda^i$. The parameter $a(\tau)$ corresponds to the familiar scale factor of the four-dimensional cosmology in our braneworld, $d\psi^2$ is an angular line element, $k=-1, 0, 1$ parametrizes the spatial curvature of the brane universe and $S_k(r)$ is given by
\begin{equation}\label{S}S_k(r) = \left\{
\begin{array}{@{}rl@{}}
\sin r & (k =1),\\
\sinh r& (k= -1),  \\[1ex]
r &(k=0). \\
\end{array}
\right .
\end{equation} 
However, the dynamics of the metric are different from the standard four-dimensional FLRW cosmology and we can show, by solving Einstein's field equations, for a given content of the universe with total energy density $\rho$ and pressure $p$, that the standard four-dimensional Friedmann equation in this modified scenario is now \cite{Deff}
\begin{equation}\label{F1}
H^2 +\frac{k}{a^2}=\bigg (\sqrt{\frac{\rho}{3M_{\rm P}^2}+ \frac{1}{r_0^2}}+\epsilon\frac{1}{2r_0^2}\bigg)^2, 
\end{equation}
where $H=\dot a/a$ is the Hubble parameter of the universe, $\epsilon=\pm 1$ and $r_0$ is the so-called crossover scale defined by $r_0 \equiv \frac{M_{\rm P}^2}{2M^3}$, where $M$ and $M_{\rm P}$ are the fundamental five- and four-dimensional Planck masses, respectively. The energy-momentum equation takes its usual form
\begin{equation}\label{F2}
\dot\rho + 3(\rho + p)H=0.
\end{equation}
Notice that we recover the standard four-dimensional Friedmann equation from (\ref{F1})
whenever $\rho /M_{\rm P}^2$ is large compared to $1/r_0^2$. 
However, at late times we may no longer assume  $\rho /M_{\rm P}^2 >> 1/r_0^2$ and the brane universe has a generically different behaviour from the ordinary four-dimensional cosmology. Depending on the sign of $\epsilon$ in the modified Friedmann equation \eqref{F1} one can identify and distinguish two distinct late time behaviours. Firstly, in the case where $\epsilon=-1$, it was shown in \cite{FE} that $a(\tau)$ diverges at late times such that the energy density $\rho$ is driven to smaller values and reaches a regime where it is small in comparison with $M_{\rm P}^2/r_0^2$. One then has a transition to a pure five-dimensional regime where the Hubble scale parameter is linear in the energy density $\rho$. This phase is referred to as the FLRW phase and will not be the subject of our present analysis.

On the other hand, if $\epsilon=+1$, we find that as the energy density crosses the threshold $M_{\rm P}^2/r_0^2$ one has a transition from the usual four-dimensional FLRW cosmology to a brane, self-inflationary solution where the Hubble scale parameter is approximately constant. Thus, the cosmology of DGP gravity provides an alternative explanation for the contemporary cosmic acceleration.
In terms of the Hubble scale, if we neglect the spatial curvature term,
$H(\tau)$ is found to evolve according to the standard Friedmann equation at early times
and when $H(\tau)$ is large, but is 
modified when $H(\tau)$ becomes comparable to $r_0^{-1}$. 
In the case of choosing a cosmological solution associated with $\epsilon=+1$ in (\ref{F1}) and assuming that the crossover scale is of the order of the Hubble radius $H_0^{-1}$, where $H_0$ is today's Hubble constant, we will obtain a scenario where
DGP gravity could explain the current cosmic acceleration, despite the fact there is no cosmological constant on the brane, through the existence of an infinite-volume extra dimension which modifies the laws of gravity at distances larger than $r_0$.
\\
The set of Eqs.(\ref{F1})-(\ref{F2}) may be used to characterise the cosmology of the DGP model and connect it to observational constraints on the real $(3+1)$-dimensional world. Therefore, we will first briefly review dark energy models in order to carry out this possible connection. 

In dark energy cosmological models, we assume that GR is still valid and that the late-time acceleration phase of the cosmic expansion is driven by a cosmological constant component with sufficient negative pressure, the so-called dark energy, interpreted as a dark fluid whose energy density is $\rho_{\rm DE}$ and pressure $p_{\rm DE}$. The content of the universe is then related to its expansion through the standard Friedmann equation. The total energy density is given by $\rho=\rho_M +\rho_{\rm DE}$, where $\rho_M$ is the energy density of cold-pressureless constituents of the universe (i.e. baryons and dark matter). 

The dark energy component may be characterised by an equation of state $w=p_{\rm DE}/\rho_{\rm DE}$, so that if $w$ is constant then by solving the energy-momentum conservation equation \eqref{F2} one obtains $\rho_{\rm DE}(\tau)=\rho^0_{\rm DE}a^{-3(1+w)}$
where $\rho_{\rm DE}^0$ is a constant. In the $\Lambda$CDM model, it is assumed that dark energy is only composed of a cosmological constant which may be interpreted as a dark energy fluid with negative pressure i.e. $w=-1$ and  $\rho_{\Lambda}(\tau)=\rho^0_{\Lambda}$.
For a constant $w$ dark energy model and assuming $H_0$ is specified, one may define the cosmological parameters $\Omega_M$, $\Omega_{\rm DE}$ and $\Omega_k$ in terms of the redshift $1+z \equiv a_0/a$ as usual:
\begin{equation}\label{omm}
\Omega_M=\Omega_M^0(1+z)^3, \quad\mathrm{where}\quad  \Omega_M^0\equiv \frac{8 \pi G \rho_M^0}{3H_0^2},
\end{equation}
\begin{equation}\label{omm2}
\Omega_{\rm DE}=\Omega_{\rm DE}^0(1+z)^{3(1+w)}, \quad\mathrm{where}\quad \Omega_{\rm DE}^0\equiv\frac{8 \pi G \rho_{\rm DE}^0}{3H_0^2},
\end{equation}
and
\begin{equation}\label{omk}
\Omega_k\equiv \frac{-k}{H_0^2a_0^2}.
\end{equation}
The Friedmann equation can be re-written as
\begin{equation}\label{Fr}
H^2(z)=H^2_0\bigg [ \Omega_k(1+z)^2+\Omega_M^0(1+z)^3+\Omega_{\rm DE}^0(1+z)^{3(1+w)}\bigg ],
\end{equation}
where in the $\Lambda$CDM model $\Omega_{\rm DE}^0(1+z)^{3(1+w)}=\Omega_{\Lambda}$.\\

Let us now consider the evolutionary behaviour of the modified Friedmann equation for DGP gravity \eqref{F1}. In this work we are concerned with the case where the universe only consists of pressureless energy-momentum constituents, $\rho_M$, while still accelerating in its late time expansion. We must then focus on the self-inflationary phase of the Friedmann equation such that $\epsilon = +1$ in (\ref{F1}). 
By analogy with the dark energy $w$-model one may introduce a new cosmological parameter $\Omega_{r_0}$
\begin{equation}\label{r_0}
\Omega_{r_0} \equiv \frac{1}{4r_0^2H_0^2},
\end{equation}
such that the self-inflationary phase of the Friedmann equation \eqref{F1} can be written as
\begin{equation}\label{H}
H^2(z)=H^2_0\bigg [ \Omega_k(1+z)^2+\bigg(\sqrt{\Omega_{r_0}}+\sqrt{\Omega_{r_0}+ \Omega_M^0(1+z)^3}\bigg)^2\bigg].
\end{equation}
Comparing this expression with the conventional Friedmann equation \eqref{F1} we notice that $\Omega_{r_0}$ mimics a dark energy parameter $\Omega_{\rm DE}^{0}$.
At the present time ($z=0$) we can find the corresponding normalisation condition for the cosmological parameters
\begin{equation}\label{norm}
\Omega_k+\bigg(\sqrt{\Omega_{r_0}}+\sqrt{\Omega_{r_0}+\Omega_M^0}\bigg)^2=1,
\end{equation} 
which is clearly different from the normalization condition of the standard four-dimensional Friedmann equation \eqref{Fr}, which requires that $\Omega_k+\Omega_M^0+\Omega_{\rm DE}^0=1$. Note also that in the case where the universe is assumed to be flat $(\Omega_k = 0)$, (\ref{norm}) gives
\begin{equation} \label{typeuni}
    \Omega_{r_{0}}=\Bigg(\frac{1-\Omega_M^0}{2}\Bigg)^2 \quad \text{and} \quad \Omega_{r_{0}} < 1.
\end{equation}

\section{Luminosity distances in higher dimensional theories with non-compact extra dimensions}
\label{sec:lumdis}

\subsection{Supernova sample}
The first dataset considered in this paper is the recent `Pantheon' type Ia supernovae (SNIe) sample \cite{Scolnic:2017caz}, which consists of 1048 objects compressed in 40 bins.  It is the largest spectroscopically confirmed supernovae sample observed to date.  With these data we can estimate the observed SNIe distance modulus $\mu$, which is related to the luminosity distance, $d_L$, as follows:
\begin{equation}\label{eq:lum}
\mu(z)= 5\log{\left[\frac{d_L (z)}{1 \text{Mpc}}\right]} +25,
\end{equation}
where $d_L$ is given in units of Mpc. In the standard statistical analysis, wherein one uses observed SNIe distance moduli to constrain cosmological model parameters, one adds to the distance modulus the nuisance parameter ${\cal M}$ -- an offset comprising the sum of the supernovae absolute magnitude and other possible systematics.  Note that ${\cal M}$ is degenerate with $H_0$, which can be taken e.g. to have a uniform prior. In the case where spatial flatness is assumed, then $d_L$ is related to the comoving distance as follows:
\begin{equation}
d_{L} (z) =\frac{c}{H_0} (1+z)D(z), \quad \quad D(z) =\frac{H_0}{c}(1+z)^{-1}10^{\frac{\mu(z)}{5}-5},
\end{equation}
where $c$ is the speed of light. 
Therefore, the normalised Hubble function $H(z)/H_0$ can be inferred by taking the inverse of the derivative of $D(z)$ with respect to the redshift using the relation 
\begin{equation}
D(z)=\int^{z}_{0} H_0 \, d\tilde{z}/H(\tilde{z}), \label{eq:dist}
\end{equation}
where $H_0$ is the Hubble parameter, considered as a prior value to normalise $D(z)$. From this point forward, it is possible to carry out a standard statistical analysis by minimizing the quantity
\begin{eqnarray} \label{eq:min}
\chi_{\mu(z)}^2
=\sum^{N}_{i=1}{\frac{\left[ \mu_{\text{obs}}(z_i) - \mu_{\text{DGP}}(z_i ; \Omega_M,\Omega_{\Lambda}, r_c)\right]^2}{\sigma^{2}_{\mu,i}}},
\end{eqnarray}
where the $\sigma^{2}_{\mu,i}$ are the measurement variances and $N$ is the number of SNIe in the total sample.

In Table \ref{tab:Pantheon} we report the values inferred for the DGP model parameters using the full Pantheon 2019 dataset. 

\begin{table}[ht]
\small
\setlength{\tabcolsep}{4.0pt} 
\setlength\abovecaptionskip{-5pt}
 \setlength\belowcaptionskip{0pt}
\begin{center}
\begin{tabular}{|c|c|c|c|c|c|}
\hline
\multicolumn{1}{|c|}{Parameter}  & $\Omega_M$ & $\Omega_r$ & $\Omega_k$ & $rH_0$& $H_0$  \\
\multicolumn{1}{|c|}{} &$0.23^{+0.10}_{-0.10}$  & $0.17^{+0.05}_{-0.05}$ & $-0.11^{+0.38}_{-0.0.36}$  & $1.20^{+0.22}_{-0.18}$& $71.53^{+0.90}_{-0.86}$ \\
\hline
\end{tabular}
\end{center}
\caption{Median values and 95\% confidence intervals for the parameters of the DGP model inferred from the Pantheon supernova sample with ${\cal M} = -19.3$.}\label{tab:Pantheon}
\end{table}

\begin{figure}[ht]
\centering
\includegraphics[width=0.6\textwidth]{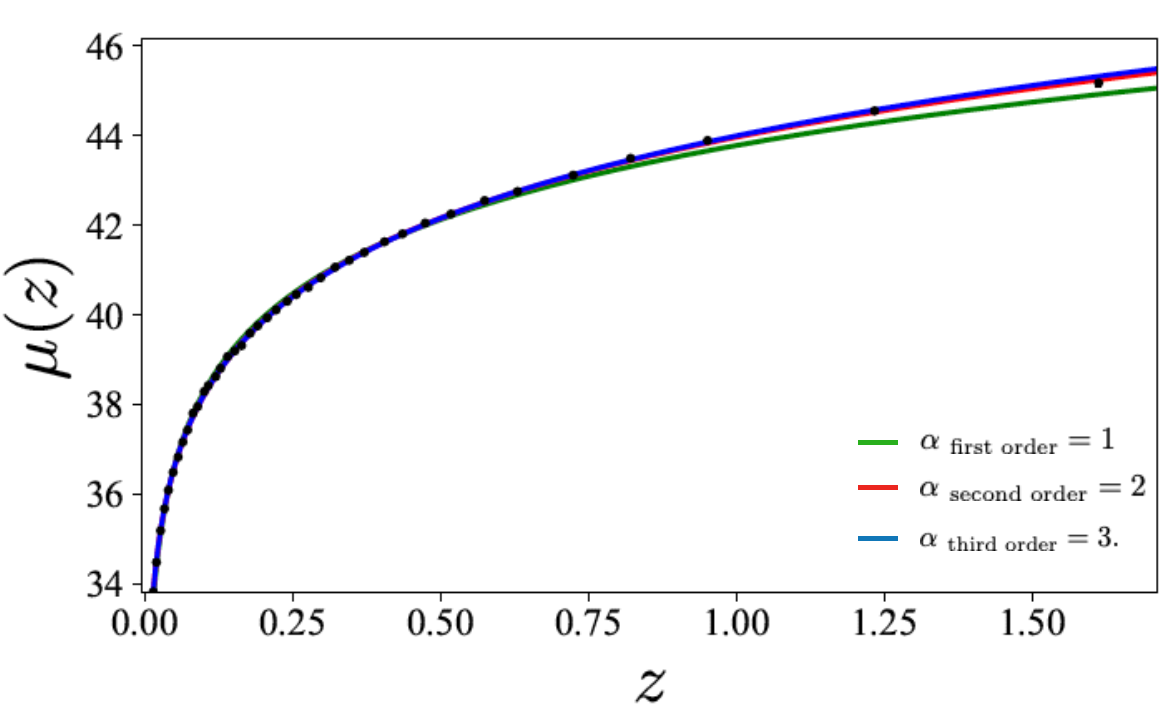}
\caption{\label{fig:pantheon_fit} Distance modulus $\mu$ as a function of $z$. The dots are the observational Pantheon sample (after subtracting the fiducial value of ${\cal M} = -19.3$). The solid lines are the reconstructed fits of $\mu(z)$ using the linear model formalism with different orders of fit.}
\end{figure}

\begin{figure}[ht]
\includegraphics[width=1\textwidth]{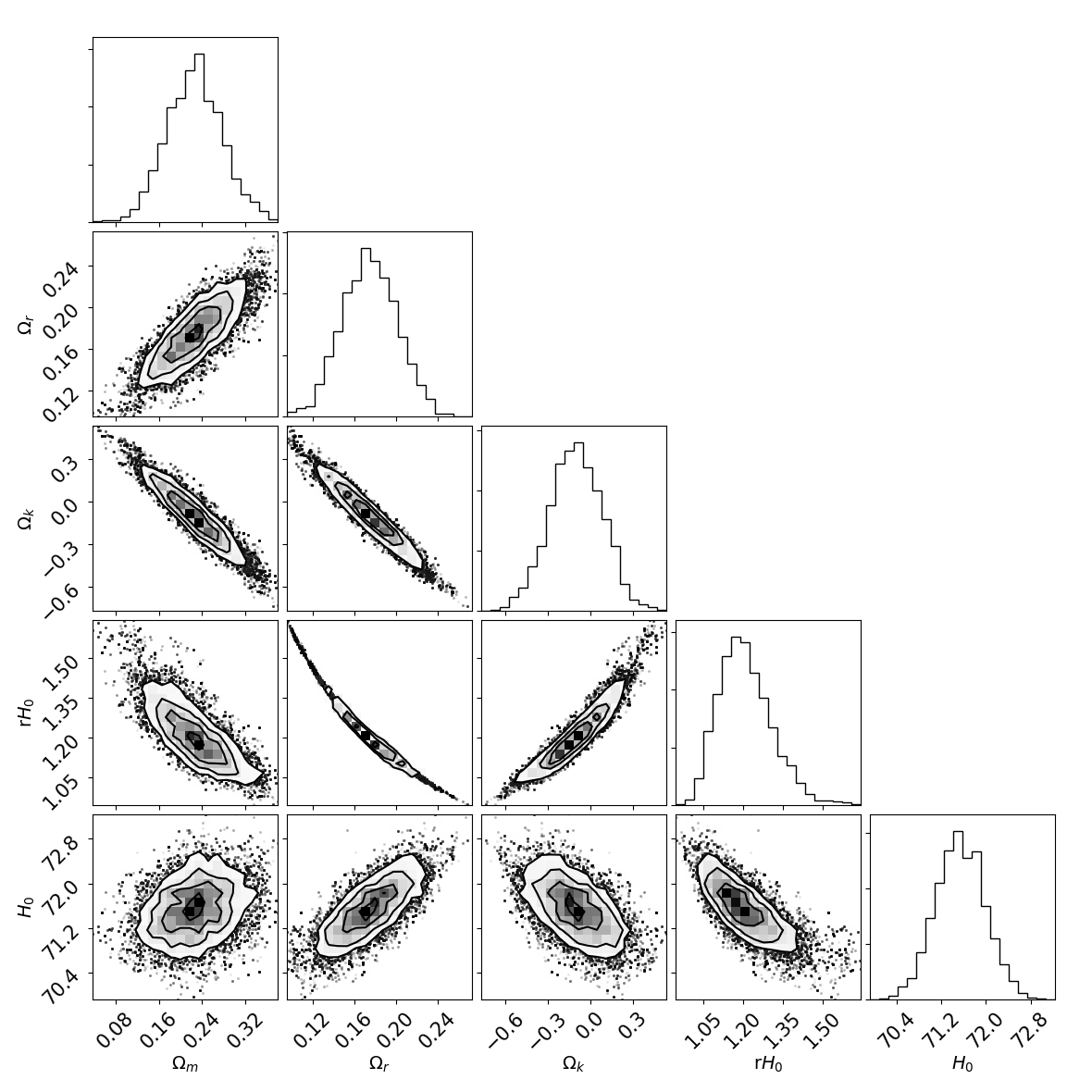}
\caption{\label{fig:pantheon_test} Joint posterior PDF for parameters of the four-dimensional DGP model parameter space, inferred from the
Pantheon supernova sample with ${\cal M} = -19.3$.}
\end{figure}

We performed a MCMC analysis using binned Pantheon data points to fit the distance moduli (having first subtracted a value of ${\cal M} = -19.3$, informed by the prior on $H_0$ from SH0ES) to a linear model at first, second and third order. The resulting fits are showed in Figure \ref{fig:pantheon_fit}. It can be seen from this Figure that a third-order model gives an excellent fit to the data.  The joint posterior on the four independent DGP model parameters is then also illustrated in Figure \ref{fig:pantheon_test}.

\subsection{Gravitational wave damping} 
In the absence of a unique GW model for higher dimensional theories with non-compact extra dimensions, we introduce a phenomenological {\em ansatz\/} for the GW amplitude scaling based on the DGP model discussed in Section \ref{section2}.  In GR the GW amplitude goes as 
\begin{equation} \label{hGR}
    h_{\rm GR} \propto \frac{1}{d_L^{\rm GW}},
\end{equation}
where $d_L^{\rm GW}$ is the luminosity distance of the GW source. Note that in GR, $d_L^{\rm GW}$ is identical to the luminosity distance $d_L^{\rm EM}$ that would be inferred from EM observations which, without loss of generality, we assume to be the `true' distance to the source.\\

Based on DGP gravity one may expect higher dimensional theories to contain a new length scale: a so-called screening scale $R_c$, beyond which they deviate from GR and exhibit gravitational leakage, leading (by flux conservation) to a reduction in the amplitude of the observed gravitational waves which now takes the form
\begin{equation} \label{hNGR}
    h_{\rm NGR} \propto \frac{1}{d_L^{\rm GW}} = \frac{1}{d_L^{\rm EM}\Bigg[1+\Big(\frac{d_L^{\rm EM}}{c R_c}\Big)^{n}\Bigg]^{(D-4)/2n}},
\end{equation}
where $D$ denotes the number of dimensions and $n$ is the transition steepness allowing for a more general \textit{shape} for the transition at the crossover scale \cite{testGR}. 
Note that (\ref{hNGR}) reduces to the standard GR scaling at distances much shorter than $R_c$ and when $D=4$ as required.

Neglecting all other effects of modified gravity on the GW amplitude, the gravitational leakage will then simply result in a measured GW distance to the source that is greater than its true luminosity distance. An event only measured in GWs would not allow us to distinguish the measured GW distance from its true distance. However, if we can independently measure $d_L^{\rm EM}$ by detecting an EM counterpart, then by comparing the two measured distances, we can place limits on the number of spacetime dimensions and the screening scale. 


\section{Forecasting LISA data: constraints on Gravitational Wave leakage} 
\label{LISA}

A gravitational observatory such as LISA will measure GWs emitted by coalescing pairs of massive black holes up to a high redshift, providing an accurate measurement of the luminosity distance to those sources.  The excellent sky localization capabilities of LISA should also greatly assist with the identification of a host galaxy or EM counterpart, thus permitting the redshift of the siren to be measured. In this section we investigate and quantify the ability of LISA to place constraints on large-scale extra-dimensional leakage of gravity from cosmologically distant standard sirens.

\subsection{MBHB synthetic catalogues} 
\label{cat}

In order to forecast LISA's ability to place limits on the number of spacetime dimensions we will make use of a number of synthetic catalogues of MBHB merger events, the gravitational radiation emitted by which would be detectable by LISA and which are considered likely to produce an EM counterpart observed by future optical and/or radio surveys -- with each merger event then being assigned a corresponding redshift, GW luminosity distance (assuming GR) and uncertainty on the luminosity distance. Such catalogues have recently been published in the context of other cosmological studies \cite{eLISA,eLISA2} and we re-employ these catalogues in the present study too. We now briefly describe the main steps involved in creating and using these synthetic catalogues, although we refer the reader to \cite{eLISA,eLISA2} for more complete technical details about how they were generated.

The first step in generating the synthetic catalogues consists of predicting the rate and redshift distribution of the population of MBHB events. In \cite{eLISA,eLISA2} these were derived using the semi-analytical model of \cite{simulation} for the evolution of the massive BHs during the formation and evolution of their host galaxies. The population was computed for three distinct scenarios regarding the initial conditions for the massive BH distribution at high redshift, namely:
\begin{enumerate}
    \item \textbf{Model popIII}: A  `realistic' light-seed scenario in which the first massive BHs are assumed to form from the remnants of population III stars (popIII) \cite{popIII1,popIII2}, and including a delay between the coalescence of MBHB host galaxies and that of the BHs themselves \cite{delays}.
    \item \textbf{Model Q3d}: A `realistic' heavy-seed scenario in which the first massive BHs are assumed to form from the collapse of protogalactic disks \cite{heavy1, heavy2, heavy3}, also including delays.
    \item \textbf{Model Q3nod}: The same model as Q3d but ignoring delays, an assumption which significantly increases the BH merger rate. This model is thus considered to be an  `optimistic' scenario for LISA's observed merger rates.
\end{enumerate}

The resulting populations of MBHB merger events for each variant MBHB model include all information about the MBHBs (masses, spins, redshift, etc.) and their host galaxies, across their entire cosmic evolution from $z \simeq 20$ to $z=0$. We refer the reader to \cite{models,simulation} for a more detailed description of the models and simulations used to generate these populations.

The next step in generating synthetic LISA catalogues consists of using the parameters of the MBHB systems modelled in each population, combined with a gravitational waveform model that permits computation of the SNR of each simulated merger event and the corresponding errors on the MBHB's waveform parameters. Again, full details are given in \cite{eLISA,eLISA2} but we provide a summary here.

For each simulated MBHB system a Fourier-domain, inspiral-only precessing waveform was first adopted, using 3.5 post-Newtonian (PN) phase evolution with 2PN spin-spin and 3.5 spin-orbit couplings, and spin-precession equations at 2PN spin-spin and 3.5PN spin-orbit orders. The SNR was calculated for each event, assuming this inspiral-only waveform, and a Fisher matrix approach was adopted to estimate the errors on the intrinsic and extrinsic parameters -- including the luminosity distance and sky location -- in each case using the LISA noise curve as described in \cite{eLISA2}. Since ultimately we want to use the sources as standard sirens we are mostly interested in the error on the luminosity distance $\Delta d_L$ and sky location $\Delta \Omega$. The reason for the latter is that the following and ultimate step is to select those events that are likely to provide a detectable optical counterpart and hence provide a measurement of the redshift to the source.

The errors on the luminosity distance and sky location estimated for this inspiral-only waveform were re-scaled to the (smaller) errors one would expect if the analysis were carried out on a full phenomC inspiral-merger-ringdown waveform, using the results presented in \cite{models} that calibrate this re-scaling. Merger events with a SNR $> 8$ and sky location error of $\Delta \Omega < 10$ $\text{deg}^2$ were then identified in \cite{eLISA,eLISA2} as detections, where $10$ $\text{deg}^2$ corresponds to the field of view of e.g. the optical Vera Rubin Observatory (VSO; previously the Large Synoptic Survey Telescope, LSST) survey \footnote{http://www.vso.org}. 

The error on the luminosity distance from the LISA measurement itself (i.e. the re-scaled error computed following calculation of the inspiral-only Fisher matrix estimate) was then combined with additional errors due to weak gravitational lensing and peculiar velocities \cite{lensing,peculiar}. Finally, the error in estimating the redshift of the host galaxy or EM counterpart was also accounted for, and the redshift error propagated through to the error budget on the luminosity distance assuming $\Lambda$CDM with Planck values for the cosmological parameters.  Two scenarios for the redshift errors were considered. (More details about the rationale for these scenarios are provided in \cite{eLISA,eLISA2}):
\begin{enumerate}
    \item{`Realistic' errors: photometric redshift measurements were assumed to have a relative 1-$\sigma$ error of $\Delta z_{\rm photo} = 0.05 (1 + z)$ while spectroscopic measurements have a relative 1-$\sigma$ error of $\Delta z_{\rm spect} = 0.01 (1 + z)^2$.}
    \item {`Optimistic' errors: photometric redshift measurements were assumed to have a relative 1-$\sigma$ error of $\Delta z_{\rm photo} = 0.03 (1 + z)$ while spectroscopic measurements have a fixed relative 1-$\sigma$ error of $\Delta \_{\rm spect} = 0.01$ and it is assumed that a `delensing' procedure can be applied that is capable of reducing by 50\% the luminosity distance error due to weak lensing.}
\end{enumerate}

In summary, the catalogues of sources presented in \cite{eLISA,eLISA2} consist of MBHBs whose gravitational radiation would be detectable by the LISA constellation \cite{LISA} with a SNR $> 8$ and $\Delta \Omega < 10$ $\text{deg}^2$ and whose EM counterpart could be visible either directly by an optical survey telescope such as the VSO or in the radio band by e.g. the Square Kilometer Array (SKA) \footnote{http://www.skatelescope.org} (with follow-up observations of the host galaxy in the optical band). As discussed in \cite{eLISA,eLISA2}, because there is a significant scatter in the characteristics of the MBHB population between different catalogues, 22 simulated four-year catalogues were considered for each MBHB formation scenario. 

\begin{figure}[ht]
  \centering
  \includegraphics[width=1.0\textwidth]{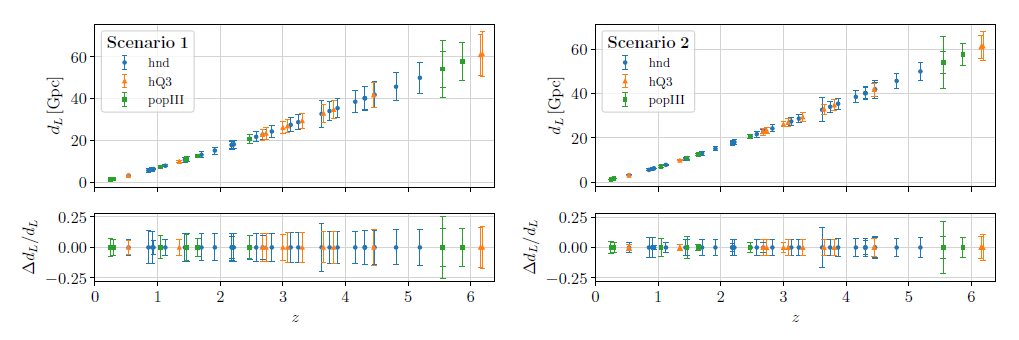}\label{LCDM}
  \setlength\abovecaptionskip{-5pt}
  \caption{Hubble diagram showing the luminosity distance as a function of redshift for one of the 22 sets of synthetic catalogues of MBHB standard sirens, assumed to be simultaneously detectable with LISA and an EM survey telescope, as presented in \cite{eLISA2} and as described in the text. The sirens in the catalogue are shown for the case of redshift errors under both Scenario 1 (i.e. `realistic') and Scenario 2 (i.e. `optimistic') as described in the text. For greater clarity the lower panels also show the fractional errors on the luminosity distance. Errors are shown for all three galaxy formation scenarios considered.  Reproduced from Figure 12 of \cite{eLISA2}.}
 \vspace{-1.em}
\label{fig:HQ3cat}
\vspace{1cm}

\end{figure}

Figure \ref{fig:HQ3cat} (reproduced from \cite{eLISA2}) shows example Hubble diagrams (upper panels) and fractional luminosity distance errors (lower panels) for one set of catalogues, for all three galaxy formation models considered, and for the case of both `realistic' and `optimistic' error scenarios, as discussed above.

Following \cite{eLISA,eLISA2}, Table \ref{statsmodels} shows for each of the MBHB formation models the average number of GW standard sirens that would be observable over a 4-year LISA mission, together with the average relative error on the luminosity distance in the `optimistic' and `realistic' scenarios described above. These averages indicate that the number of `useful' standard sirens is substantially larger for the Q3nod model, thereby potentially improving the precision with which the parameters of our cosmological model can be inferred. The other two galaxy formation models are expected to produce very similar numbers of observable sources, with the popIII model being slightly worse.

\begin{table}[h]
\vspace{-0.3em}
\setlength\abovecaptionskip{-2pt}
 \setlength\belowcaptionskip{2pt}
\begin{center}
\begin{tabular}{*{5}{c c c c}}
\hline                 
Model&Total number&$\Delta d_L$ (optimistic)&$\Delta d_L$ (realistic)\cr 
\hline
popIII & $13.6$ & $0.0929$ & $0.139$ \cr \rule{0pt}{4ex}
Q3d & $14.7$ & $0.0645$ & $0.122$\cr  \rule{0pt}{4ex}
Q3nod & $28.3$ & $0.0660$ & $0.119$ \cr
\hline
\end{tabular}
\end{center}
\caption{\label{statsmodels}  Key numbers summarising the synthetic four-year catalogues from \cite{eLISA,eLISA2}, for all three MBHB formation models. The second column gives the average number of LISA detections with an EM counterpart, assuming a 4-year mission. The last two columns present the average relative error $\Delta d_L$ on luminosity distance, for the case of `optimistic' and `realistic' errors respectively.}
\end{table}

\subsection{Generating GW data} 
\label{datamodel}

Equipped with the synthetic catalogues of standard sirens generated for $\Lambda$CDM cosmology, and with the required methodology (see Eq. \ref{hNGR}) to describe the effect of gravitational damping on the GW waveform in higher dimensional theories with non-compact extra dimensions, the next step is to generate realisations of MBHB sirens for various possible higher-dimensional theories. 

Since we always assume that light and matter is restricted to the four-dimensional universe, we first infer the `true' EM luminosity distance $(d_L^{\rm EM})_i$ for each MBHB event of each catalogue by plugging its corresponding redshift $z_i$ into the usual distance-redshift relation \cite{Distances}. 
Note that $d_L^{\rm EM}$ depends on the value of the Hubble constant (which, with loss of generality, we assume to be equal to the SH0ES value of $H_0 =73.24\pm 1.74$ km $\text{s}^{-1}$ $\text{Mpc}^{-1}$ \cite{Shoes}, and on the assumed value of the cosmological parameters $\Omega_M^0$ and $\Omega_\Lambda$, which we fit using the Pantheon supernovae sample.

We note, of course, that to be fully self-consistent, then we should compute the MBHB merger rates and redshift distributions in our particular chosen higher dimensional theory and not in $\Lambda$CDM.  However, we do not expect the rates and distributions obtained in that manner to be significantly different from those obtained assuming $\Lambda$CDM, since the dominant effect is instead the details of the galaxy formation and evolution model adopted.  Thus, for simplicity we will continue to adopt the merger rates and redshift distribution in each formation model previously calculated for the $\Lambda$CDM case. 

For the $i^{\rm th}$ siren we then compute the corresponding `true' GW luminosity distance $(d_{\rm GW})_i \equiv d_L^{\rm GW}\{(d_L^{\rm EM})_i; D,R_c,n\}$, by inverting the scaling relation (\ref{hNGR}). The GW distance depends not only on the `true' EM distance, but also on the cosmological parameters of the higher-dimensional gravity. In what follows we consider $D= 5, 6$ or $7$, $n=1$ or $10$ and $R_c=R_H \sim 4$ Gpc or $4R_H$ where $R_H=H_0^{-1}$ is the current Hubble radius. The reason for these particular choices of screening scale is that, based on the DGP model, one may expect the characteristic length scale for higher-dimensional theories of gravity to be of the order of the Hubble radius. The values for the steepness parameter on the other hand were chosen arbitrarily but such that they have not yet been ruled out by the comparison of distance measurements from GW and EM observations of GW170817 \cite{testGR}. 

Finally, for the $i^{\rm th}$ siren we generate a `measured' GW luminosity distance $(x_{\rm GW})_i$ by adding stationary Gaussian noise with zero mean and standard deviation given by $\sigma_i=(\Delta d_L)_i \times (d_L^{\rm GW})_i$, where $(\Delta d_L)_i$ is the relative error on the corresponding EM distance, calculated in either the `optimistic' or `realistic' scenario.  This approach mirrors that adopted in \cite{eLISA,eLISA2}.  

We note that the addition of Gaussian scatter to the GW luminosity distance could in principle result in MBHBs, particularly those at highest redshift in the simulated catalogues, dropping below the adopted SNR threshold for observable events. We have investigated this possible selection bias by repeating the analyses described in the following sections but taking the conservative step of eliminating the most distant sirens at $z \sim 8$.  We find that the effect of this sample correction on our parameter estimation is small: it has minimal impact on the mean estimated parameters and results in only a slight increase of their variance -- and the effect is much smaller than the variation between the different galaxy formation models.  We therefore do not consider this possible selection bias any further in this paper.

\section{Statistical analysis methods} 
\label{bayes}
Given two sets of $j$ statistically independent GW luminosity distance and EM redshift measurements for MBHB merger events from a given catalogue $C$, ${{\boldsymbol x}^C_{\rm GW}} = \left\{(x_{\rm GW})_1,(x_{\rm GW})_2...,(x_{\rm GW})_j)\right\}$ and ${\boldsymbol{z}}^C=\left\{z_1, z_2,...,z_j)\right\}$ respectively, we carried out a Bayesian analysis to infer the posterior of the number of spacetime dimensions $D$, the screening scale $R_c$, the steepness parameter $n$ and Hubble constant $H_0$.
We then carried out a Bayesian model selection analysis to evaluate the evidence for, or against, each higher-dimensional model by considering the full analyses in comparison to DGP scenarios as pivot models. These results allow us to assess the capability of LISA to place constraints on higher-dimensional theories of gravity.

We first briefly describe the statistical framework of Bayesian parameter estimation, before then outlining the Bayesian model selection technique used.

\subsection{Parameter Estimation}
The joint posterior of the parameters $\boldsymbol{\theta}^C= \{ D, R_c, n, H_0 \}$ for a given catalogue realisation, $C$, is obtained by applying Bayes' theorem \cite{Bayes}
\begin{equation}\label{Bays}
p({\boldsymbol{\theta}}^C | {\boldsymbol{z}}^C,{\boldsymbol{x}}_{\rm GW}^C, {\bf M}) = \frac{p({\boldsymbol{z}}^C,{\boldsymbol{x}}_{\rm GW}^C | {\boldsymbol{\theta}}^C, {\bf M}) p({\boldsymbol{\theta}}^C | {\bf M})}{Z} ,
\end{equation}
where $p({\boldsymbol{z}}^C,{\boldsymbol{x}}_{\rm GW}^C | {\boldsymbol{\theta}}^C)$ is the joint likelihood for GW and EM data from the catalogue realisation, $p({\boldsymbol{\theta}}^C | {\bf M})$ is the prior probability distribution, {\bf M} represents all the background information that has gone into defining our model and $Z=p({\boldsymbol{z}}^C,{\boldsymbol{x}}_{\rm GW}^C | {\bf M})$ is a normalisation constant -- the so-called evidence, obtained by multiplying the likelihood by the prior and integrating over all parameters that define the model. We will make use of $Z$ when comparing the models in (\ref{odds}) but since $Z$ does not depend on $\boldsymbol{\theta}^C$, it is a constant for a given model and the posterior PDF is just proportional to the product of the prior distribution on the parameters and the likelihood of the data given the parameters.

\begin{figure}[ht]
\begin{center}
\includegraphics[width=0.8\textwidth]{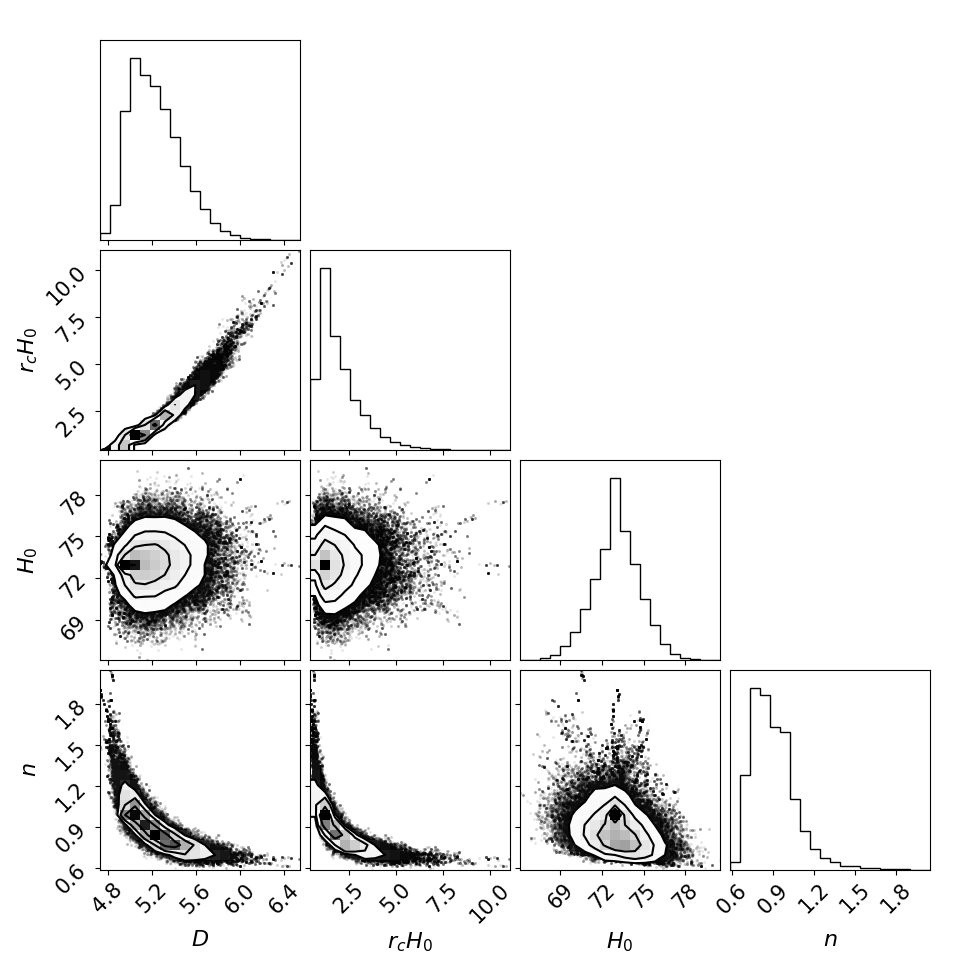}
\caption{\label{fig:corner16_1} Joint posterior PDF for the full, four-dimensional parameter space for a randomly chosen catalogue within the Q3d formation model and assuming optimistic errors. The true model parameters for this example are ${\boldsymbol{\theta}} = \{ D=5,R_c=R_H,n=1,H_0 \}$ . On the axes are superimposed the one-dimensional marginalised distributions. The two-dimensional plots show the contours at 0.5, 1, 1.5, and 2-$\sigma$.} 
\end{center}
\end{figure}

We can write the joint likelihood for the GW data ${\boldsymbol{x}}_{\rm GW}^C$ and EM observables ${\boldsymbol{z}}^C$ given ${\boldsymbol{\theta}}^C = \{ D, R_c, n, H_0 \}$ as
\begin{equation}\label{likelihood} 
p({\boldsymbol{z}}^C,{\boldsymbol{x}}_{\rm GW}^C | {\boldsymbol{\theta}}^C, {\bf M})  \propto \exp\sum_{i=1}^{j}\bigg[ -\frac{\left|(x_{GW})_i-(d_L^{GW})_i\right|^2}{2\sigma_i^2}\bigg ],
\end{equation}
where the sum runs over all MBHB events in the catalogue.

In order to represent complete ignorance about the parameters defining the higher-dimensional theory we take uniform uninformative priors in the range $D \in [3,11]$, $n \in [0,100]$ and $R_c \in [20,\infty)$ where the lower limit on the screening scale is set by distances ruled out by GW170817 \cite{pardo} and the ranges on $D$ and $n$ were chosen to limit the computational cost of the parameter estimation method. 
For the prior on the Hubble constant we consider the SH0ES measurement and adopt a Gaussian prior centred on $H_0=73.24 \, \text{kms}^{-1}\text{Mpc}^{-1}$ with standard deviation $\sigma_{H_0}=1.74 \, \text{kms}^{-1}\text{Mpc}^{-1}$. 
The prior distribution on the parameter vector ${\boldsymbol{\theta}}^C$ is then 
\begin{equation}
p({\boldsymbol{\theta}}^C | {\bf M})=p(D) p(R_c) p(n) p(H_0) \propto \mathcal{N}({H_0},\sigma_{H_0}).
\end{equation}
Results for parameters of interest are found by marginalising the joint posterior over any unwanted parameters, for example
\begin{equation}\label{pdf}
p(\theta_{1} | {\boldsymbol{x}}_{\rm GW}^C, {\boldsymbol{z}}^C, {\bf M}) = \int{d\theta_{2}...d\theta_{m} 
\, \, p({\boldsymbol{\theta}}^C | 
{\boldsymbol{x}}_{\rm GW}^C, {\boldsymbol{z}}^C, {\bf M})}	.
\end{equation}
The marginalised posterior PDF of a given parameter can then be used to find its median and construct credible regions. 

As can be seen from (\ref{pdf}) computing the marginalised PDFs requires the evaluation of multi-dimensional integrals which is computationally intensive for the size of the parameter space and the amount of data to consider. This is addressed by using a stochastic sampling engine based on the MCMC algorithm which samples random draws from the target posterior distributions allowing us to approximate the desired integrals.\footnote{In this analysis we used the python package emcee by Foreman-Mackey et al.(2013) \url{http://dan.iel.fm/emcee}, which implements the affine-invariant ensemble sampler of Goodman \& Weare (2010) \url{https://github.com/grinsted/gwmcmc} to perform the MCMC algorithm.}
The results are presented in Sec. \ref{resultsLISA}.

\subsection{Model selection}
\label{odds}

To further quantify the ability of future LISA observations to place constraints on modified gravity theories -- or extra-dimensional theories -- we want to compute the probability that, for data simulated for a particular non-GR (denoted NGR)\footnote{In this paper the NGR models are taken to be those scenarios with $D \geq 6$.} model, ${\bf M}_{\rm NGR}$, characterised by the parameters ${\boldsymbol{\theta}}_{\rm NGR}=\{D,R_c,n,H_0\}$, the data will favour the NGR model over the nested DGP model, ${\bf M}_{\rm DGP}$, i.e. the pivot model parametrised by ${\boldsymbol{\theta}}_{\rm DGP} = \{D=5,R_cH_0=1.2,n,H_0\}$. This step is achieved by computing the posterior odds ratio, defined as the ratio of the posterior probabilities of the two competing models:
\begin{equation}
    O_{\rm NGR/GR}=\frac{p({\bf M}_{\rm NGR}) \, p({\boldsymbol{z}},{\boldsymbol{x}}_{\rm GW} | {\bf M}_{\rm NGR}) }{p({\bf M}_{\rm DGP}) \, p({\boldsymbol{z}},{\boldsymbol{x}}_{\rm GW} | {\bf M}_{\rm DGP})} \, = \, \frac{p({\bf M}_{\rm NGR})}{p({\bf M}_{\rm DGP})} \, B_{\rm NGR/DGP}, \quad
\end{equation}
which is simply the prior odds multiplied by the ratio of the evidence for each model -- the so-called Bayes factor in favour of ${\bf M}_{\rm NGR}$. 
If we further assume  that the two models are equally probable a priori, such that $p({\bf M}_{\rm NGR})=p({\bf M}_{\rm DGP})$, then the posterior odds is simply equal to the Bayes factor which from Bayes' theorem considered above (\ref{Bays}) may be now expressed as
\begin{equation}
    \label{bayesfact}
    B_{\rm NGR/DGP}=\frac{Z_{\rm NGR}}{Z_{\rm DGP}}=
    \frac{ 
    \int{ d {\boldsymbol{\theta}}_{\rm NGR}} \, p({\boldsymbol{\theta}}_{\rm NGR} | {\bf M}_{\rm NGR}) \, 
    p({\boldsymbol{z}},{\boldsymbol{x}}_{\rm GW} | {\boldsymbol{\theta}}_{\rm NGR}, {\bf M}_{\rm NGR})}{
    \int{ d {\boldsymbol{\theta}}_{\rm DGP}} \, p({\boldsymbol{\theta}}_{\rm DGP} | {\bf M}_{\rm DGP}) \, 
    p({\boldsymbol{z}},{\boldsymbol{x}}_{\rm GW} | {\boldsymbol{\theta}}_{\rm DGP}, {\bf M}_{\rm DGP})} .
\end{equation}
According to Jeffrey's scale: if $B_{\rm NGR/DGP}$ is larger (smaller) than unity then this tells us that ${\bf M}_{\rm NGR}$ is more (less) strongly supported by the data under consideration than ${\bf M}_{\rm DGP}$.

\section{Results and discussion} \label{resultsLISA}
The MCMC implementations described above were run on the simulated GW and EM data for each of the 22 catalogues describing the three MBHB formation scenarios and for each true cosmological model presented in Sec.(\ref{datamodel}). We first discuss the results of the parameter estimation, then consider our model comparison results.

\subsection{Parameter estimation using MCMC}
 \label{resultsMCMC}
To quantify the capability of LISA to constrain a particular cosmological parameter for a given cosmological scenario and MBHB formation model, we first record the median values of the marginalised PDFs of each cosmological parameter for all the catalogue realisations. We then use the median as a Figure-of-Merit (FoM) for these parameter estimates.  For an estimate of the LISA error on the median FoM we first take the $95\%$ credible interval (CI) around the median value for each catalogue and then pick the median of these intervals to represent the $95\%$ CI of the FoM. 

There is also a significant scatter in the characteristics of the MBHB population between different catalogues. To gain an idea of the impact of this scatter on the cosmological constraints that we can place, we later show box plots representing the distributions of these constraints for each true cosmological model and MBHB formation model considered. Note that when we compare the capability of LISA to place constraints on a given parameter for different cosmologies and MBHB models, we will always use the median FoM together with $95\%$ CI, which essentially captures the scatter in the FoM and hence provides a realistic estimate of the statistical uncertainty that can be expected.

Figure \ref{fig:corner16_1} shows an example of the joint posterior PDF over our four-dimensional model parameter space (i.e.$\, \, \boldsymbol{\theta}^C = \{ D, R_c, n, H_0 \}$) for a randomly chosen catalogue, and for a representative true cosmological scenario and MBHB formation model, assuming optimistic errors.

Tables \ref{PEstatsD}, \ref{PEstatsR}, \ref{PEn} and \ref{PEH0} present FoMs, derived from the marginalised PDFs, for each of the cosmological model parameters ${\boldsymbol{\theta}} = \{ D,R_c,n,H_0 \}$ respectively, for all MBHB formation scenarios and for the range of true cosmological models considered, in the case of both optimistic and realistic errors on the siren luminosity distance estimates. The results for the steepness parameter and Hubble constant will not be considered in detail as they are not of astrophysical interest to our analysis and do not add any information about LISA's ability to constrain modified gravity that we cannot infer by considering the results for $D$ and $R_c$. 

In each entry of Tables \ref{PEstatsD}, \ref{PEstatsR}, \ref{PEn} and \ref{PEH0}, the top row shows the median FoM and $95\%$ CI for light seeds (popIII), the central row for heavy seeds with delays (Q3d) and the bottom row for heavy seeds without delays (Q3nod). Q3nod systematically gives better results than the other two scenarios, which are roughly comparable to each other, due to their lower number of detectable standard sirens, as shown in Table \ref{statsmodels}. Clearly, this demonstrates that the extent to which LISA can be used to perform meaningful constraints on theories of modified gravity defined by the scaling $\eqref{hNGR}$ will strongly depend on the actual redshift distribution of MBHB merger events and the corresponding efficiency in identifying an EM counterpart. In the following, whenever we need to restrict to one MBHB formation model, we always choose Q3d, since it is the intermediate scenario (as far as the number of standard sirens is concerned) among those we consider. The constraints would be comparable or slightly worse for popIII and somewhat better for the model Q3d. Also whenever we need to show likelihood contours we restrict to the contour of the catalogue representing the FoM.\\

\renewcommand{\arraystretch}{1.7}
\begin{table}[ht]
\begin{center}
\small
\setlength{\tabcolsep}{4.0pt} 
\setlength\abovecaptionskip{-5pt}
 \setlength\belowcaptionskip{0pt}
\caption{\label{PEstatsD} Median figures of merit and $95\%$ credible intervals summarising the marginalised posterior PDF of $D$, the number of dimensions, for the various assumed true cosmological scenarios and MBHB formation models considered. In each row of the Table separated by dashed lines, the top sub-row shows the FoMs and CI for light-seeds (popIII), the central sub-row for heavy seeds with delays (Q3d) and the bottom sub-row for heavy seeds without delays (Q3nod).  Results in the top hand panel assume optimistic distance errors for the MBHB sirens, while those in the bottom hand panel assume realistic distance errors.
}
\vspace{4mm}

\begin{tabular}{|c|c|c||c|c|}
\hline 
\multicolumn{1}{|c|}{ } & \multicolumn{4}{c|}{Inferred $D$ (optimistic errors)}  \\
\hline
\multirow{2}{*}{Model} & \multicolumn{2}{c||}{$R_c=R_H$} & 
    \multicolumn{2}{c|}{$R_c=4R_H$}  \\
\cline{2-5}
 & $n=1$ & $n=10$ & $n=1$ & $n=10$  \\
\hline
 \multirow{3}{*}{$D=5$}& $5.02^{+0.60}_{-0.28}$  & $5.03^{+0.12}_{-0.09}$&$5.08^{+0.92}_{-0.59}$ & $5.00^{+1.66}_{-0.33}$  \\
 \cdashline{2-5}
 & $5.00^{+0.55}_{-0.24}$  & $5.01^{+0.10}_{-0.08}$&$5.18^{+0.85}_{-0.55}$ &$5.05^{+0.69}_{-0.26}$  \\
\cdashline{2-5}
& $5.03^{+0.41}_{-0.22}$  & $5.01^{+0.06}_{-0.06}$&$5.16^{+0.77}_{-0.50}$ &$4.98^{+0.39}_{-0.21}$   \\
\hline
\multirow{3}{*}{$D=6$}& $5.97^{+0.59}_{-0.35}$  & $6.02^{+0.11}_{-0.09}$&$6.18^{+1.39}_{-0.90}$ &$6.08^{+0.59}_{-0.38}$  \\
\cdashline{2-5}
 & $5.98^{+0.50}_{-0.33}$  & $6.00^{+0.10}_{-0.08}$&$6.21^{+1.34}_{-0.75}$ &$6.03^{+0.39}_{-0.27}$  \\
\cdashline{2-5}
&  $6.05^{+0.46}_{-0.30}$  & $6.00^{+0.07}_{-0.06}$&$6.18^{+1.24}_{-0.67}$ &$6.02^{+0.26}_{-0.22}$ \\
\hline 
\multirow{3}{*}{$D=7$}& $7.09^{+0.68}_{-0.45}$  & $7.00^{+0.11}_{-0.10}$&$7.48^{+2.05}_{-1.14}$ &$7.03^{+0.51}_{-0.37}$   \\
\cdashline{2-5}
 & $6.99^{+0.53}_{-0.38}$  & $7.00^{+0.10}_{-0.08}$&$7.18^{+1.63}_{-0.88}$ &$7.00^{+0.30}_{-0.25}$\\
\cdashline{2-5}
& $7.03^{+0.46}_{-0.31}$  & $7.00^{+0.06}_{-0.06}$&$7.25^{+1.46}_{-0.81}$ &$7.03^{+0.26}_{-0.21}$ \\
\hline
\end{tabular}

\begin{tabular}{|c|c|c||c|c|}
\hline 
\multicolumn{1}{|c|}{ } & \multicolumn{4}{c|}{Inferred $D$ (realistic errors)}  \\
\hline
\multirow{2}{*}{Model} & \multicolumn{2}{c||}{$R_c=R_H$} & 
    \multicolumn{2}{c|}{$R_c=4R_H$}  \\
\cline{2-5}
 & $n=1$ & $n=10$ & $n=1$ & $n=10$  \\
\hline
 \multirow{3}{*}{$D=5$}& $5.05^{+1.38}_{-0.40}$  & $5.01^{+0.27}_{-0.19}$&$5.25^{+1.53}_{-0.76}$ & $5.41^{+3.61}_{-0.73}$  \\
 \cdashline{2-5}
 & $4.98^{+1.00}_{-0.29}$  & $5.07^{+0.39}_{-0.19}$&$5.10^{+1.18}_{-0.57}$ &$5.28^{+2.45}_{-0.66}$  \\
\cdashline{2-5}
& $5.00^{+0.80}_{-0.33}$  & $5.00^{+0.15}_{-0.13}$&$5.17^{+1.01}_{-0.64}$ &$5.17^{+1.77}_{-0.43}$   \\
\hline
\multirow{3}{*}{$D=6$}& $5.97^{+1.16}_{-0.55}$  & $6.02^{+0.24}_{-0.20}$&$6.30^{+1.71}_{-1.22}$ &$6.32^{+3.50}_{-0.87}$  \\
\cdashline{2-5}
 & $6.00^{+1.07}_{-0.49}$  & $5.99^{+0.24}_{-0.19}$&$6.26^{+1.60}_{-1.07}$ &$6.06^{+0.86}_{-0.45}$  \\
\cdashline{2-5}
&  $6.05^{+0.82}_{-0.43}$  & $6.01^{+0.15}_{-0.13}$&$6.43^{+1.46}_{-1.08}$ &$5.98^{+0.68}_{-0.39}$ \\
\hline 
\multirow{3}{*}{$D=7$}& $7.09^{+1.35}_{-0.72}$  & $7.04^{+0.28}_{-0.22}$&$7.60^{+2.31}_{-1.64}$ &$7.24^{+1.76}_{-0.77}$   \\
\cdashline{2-5}
 & $7.05^{+1.13}_{-0.58}$  & $7.02^{+0.21}_{-0.16}$&$7.24^{+2.10}_{-1.29}$ &$7.04^{+0.72}_{-0.50}$\\
\cdashline{2-5}
& $7.00^{+0.86}_{-0.53}$  & $7.00^{+0.14}_{-0.12}$&$7.62^{+1.92}_{-1.37}$ &$7.02^{+0.53}_{-0.40}$ \\
\hline
\end{tabular}
\vspace{8mm}
\end{center} 
\end{table}

\begin{figure}[ht] 
    \captionsetup[subfloat]{farskip=0.2pt,captionskip=1.pt}
     \centering
     \subfloat{%
       \includegraphics[scale=.33]{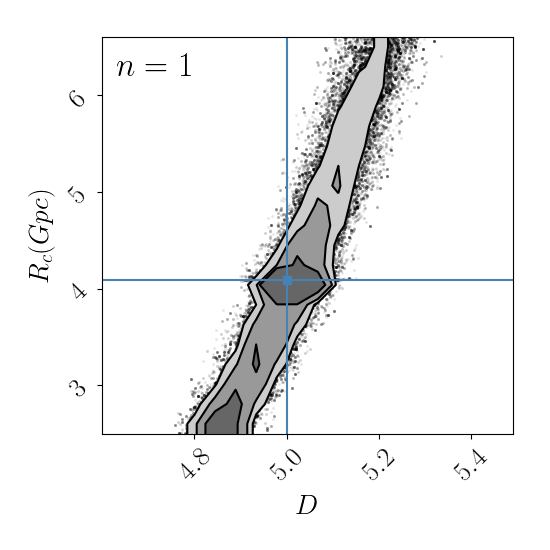}
     }%
     \subfloat{%
       \includegraphics[scale=.33]{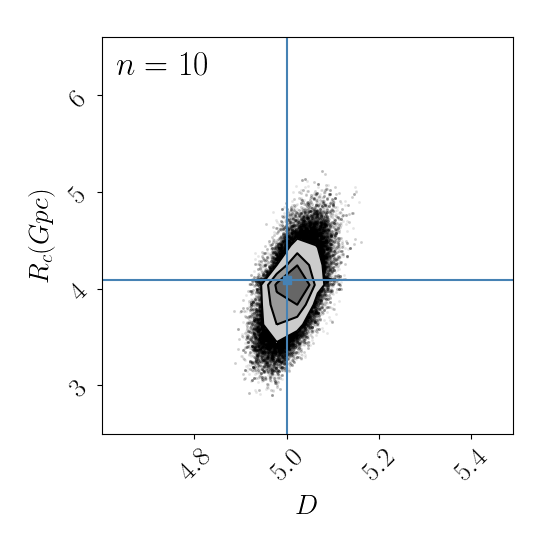}
     }%
      \subfloat{%
       \includegraphics[scale=.33]{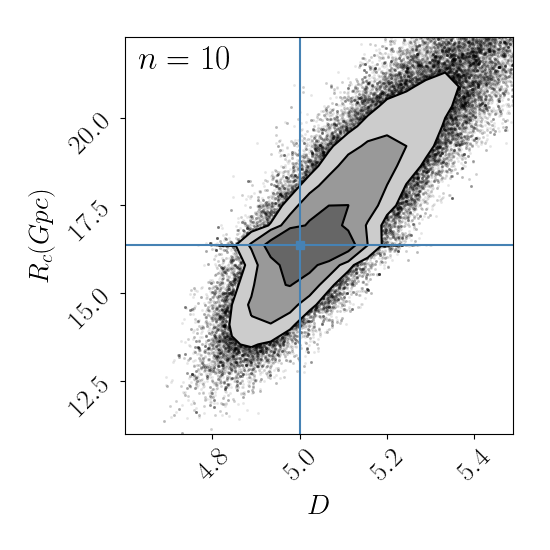}
     }\\ 
     \subfloat{%
       \includegraphics[scale=0.33]{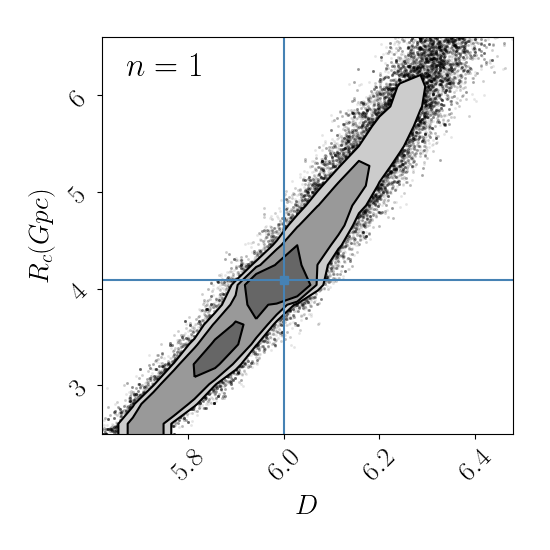}
     }%
     \subfloat{%
       \includegraphics[scale=.33]{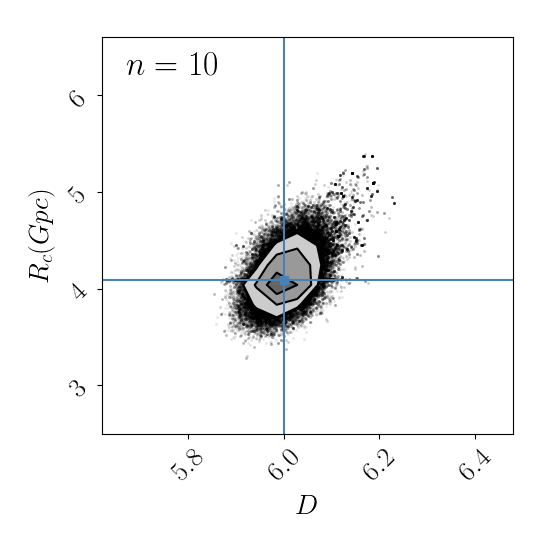}
     }
      \subfloat{%
       \includegraphics[scale=.33]{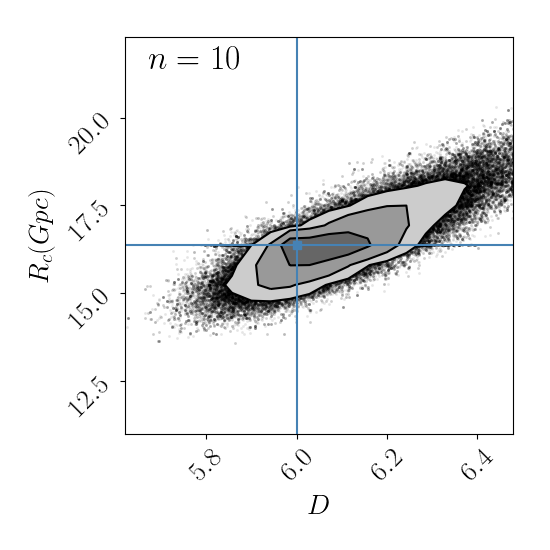}
     }\\ 
     \subfloat{%
       \includegraphics[scale=0.33]{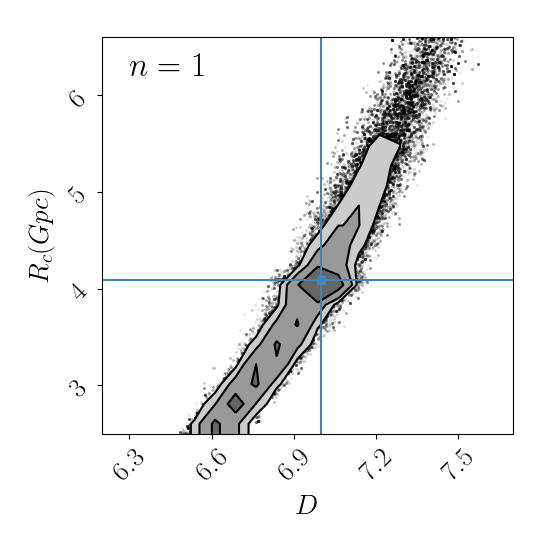}
     }%
     \subfloat {%
       \includegraphics[scale=.33]{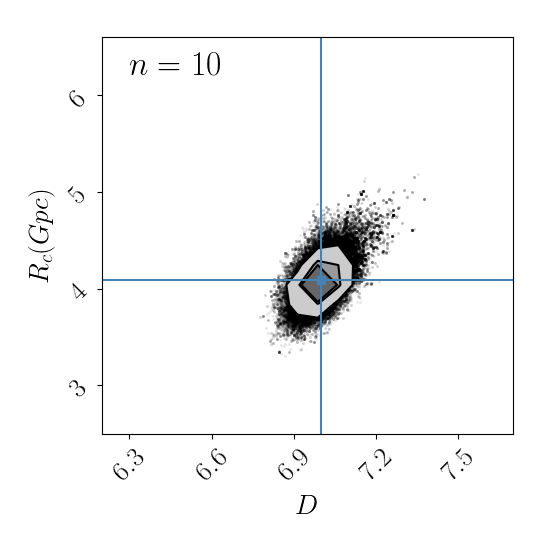}
     }%
      \subfloat{%
       \includegraphics[scale=.33]{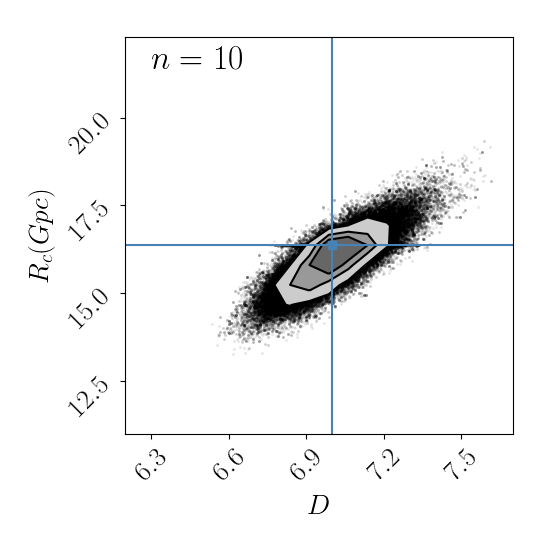}
      }%
     \caption{Marginalised contours, at 0.5, 1, 1.5, and 2-$\sigma$, for $D$ and $R_c$ in the heavy seeds (Q3d) model assuming optimistic errors. Results shown are for the following true cosmological scenarios: from left to right $(n=1, R_c=H_d)$,  $(n=10, R_c=H_d)$ and $(n=10, R_c=4H_d)$ and from top to bottom $D= 5, 6, 7$. The blue cross hairs indicate the true values of the parameters. }
     \label{fig:contDn}
   \end{figure}


\subsection{Constraints on the number of spacetime dimensions} 
We first consider LISA's ability to place limits on the number of spacetime dimensions. The FoMs and $95\%$ CI for all MBHB formation models can be found in Table \ref{PEstatsD}. We see that, except for the $R_c= 4R_H, n=1$ model, the estimates of the number of dimensions are very close to their true values for all MBHB models, indicating that LISA should be able to place useful limits on the number of dimensions in this case. Note that although the FoMs are very close to their true values for all MBHB models, the spread around the FoMs is narrower for the Q3nod model -- giving constraints on $D$, for the most favorable cosmological model $(D=7,n=10, R_c=R_H)$, at the level of $0.86 \%$ compared to $1.43\%$ and $1.57\%$ for the Q3d and popIII formation models respectively.

Furthermore, we can see from Table \ref{PEstatsD} that the steeper the transition and the smaller the screening scale the more accurate the estimates on $D$ and the smaller the confidence interval; this is why LISA will not be able to place meaningful constraints on $D$ in the case where $R_c= 4R_H$ and $n=1$.

In Figure \ref{fig:contDn} we show the likelihood contours for $D$ and $R_c$, for true values of $D=5, 6, 7$ when $n=1,R_c= R_H$; $n=10,R_c= R_H$ and $n=10,R_c= 4R_H$, for representative catalogues in the heavy seeds (Q3d) MBHB formation scenario. This Figure allows us to understand the impact of the steepness parameter and/or screening scale on the constraints on $D$. This dependency on the steepness parameter and screening scale is not surprising since one would expect that the less pronounced the modification from GR at `small' redshifts, the more distant the sirens must be for a difference in the physics to be noticeable.

As a final note to the constraints that can be placed on $D$, Table \ref{PEstatsD} shows that the precision with which we can estimate $D$ becomes marginally better as the number of dimensions increases, going from $1.2 \%$ to $0.86 \%$ in the most favourable cosmological scenario, but is not very sensitive to the true value of $D$. 

\begin{table}[ht]
\begin{center}
\small
\setlength{\tabcolsep}{4.0pt} 
\setlength\abovecaptionskip{-5pt}
 \setlength\belowcaptionskip{0pt}
\caption{\label{PEstatsR} 
Median figures of merit and $95\%$ credible intervals summarising the marginalised posterior PDF of the screening scale, $R_c$,for the various assumed true cosmological scenarios and MBHB formation models considered, for both optimistic (top panel) and realistic (bottom panel) errors. The Hubble radius is $R_H=4.093$ Gpc. 
}
\vspace{4mm}

\begin{tabular}{|c|c|c||c|c|}
\hline 
\multicolumn{1}{|c|}{ }  & \multicolumn{4}{c|}{Inferred $R_c$ (optimistic errors)} \\
\hline
\multirow{2}{*}{Model} & \multicolumn{2}{c||}{$R_c=R_H$} & \multicolumn{2}{c|}{$R_c=4R_H$} \\
\cline{2-5}
 & $n=1$ & $n=10$ & $n=1$ & $n=10$  \\
\hline
 \multirow{3}{*}{$D=5$}&  $1.03^{+2.93}_{-0.61}$  & $1.02^{+0.24}_{-0.17}$&$5.76^{+8.74}_{-4.65}$ & $4.00^{+5.44}_{-1.15}$ \\
 \cdashline{2-5}
  & $1.06^{+1.75}_{-0.50}$  & $1.02^{+0.15}_{-0.14}$&$5.46^{+8.56}_{-3.70}$ & $3.99^{+1.11}_{-0.69}$ \\
\cdashline{2-5}
& $1.06^{+2.57}_{-0.57}$  & $1.03^{+0.21}_{-0.17}$&$6.10^{+8.37}_{-4.39}$ & $4.19^{+2.90}_{- 0.96}$   \\
\hline
\multirow{3}{*}{$D=6$}&  $0.96^{+1.02}_{-0.41}$  & $1.01^{+0.12}_{-0.11}$&$4.92^{+8.32}_{-3.12}$ & $4.05^{+0.91}_{-0.58}$ \\
\cdashline{2-5}
 & $1.09^{+0.91}_{-0.41}$  & $1.00^{+0.09}_{-0.09}$&$4.95^{+7.93}_{-2.77}$ & $4.06^{+0.43}_{-0.36}$  \\
\cdashline{2-5}
 & $0.97^{+0.96}_{-0.39}$  & $1.01^{+0.13}_{-0.10}$&$4.99^{+8.20}_{-3.09}$ & $4.06^{+0.68}_{-0.47}$ \\
\hline 
\multirow{3}{*}{$D=7$}&  $1.08^{+0.78}_{-0.37}$  & $1.01^{+0.10}_{-0.08}$&$5.50^{+7.91}_{-2.98}$ & $4.06^{+0.54}_{-0.38}$  \\
\cdashline{2-5}
 & $0.99^{+0.45}_{-0.27}$  & $1.00^{+0.08}_{-0.07}$&$4.56^{+5.36}_{-2.14}$ & $4.04^{+0.29}_{-0.24}$ \\
\cdashline{2-5}
& $0.99^{+0.66}_{-0.34}$  & $1.00^{+0.10}_{-0.08}$&$4.54^{+6.63}_{-2.30}$ & $4.02^{+0.38}_{-0.33}$\\
\hline
\end{tabular}

\begin{tabular}{|c|c|c||c|c|}
\hline 
\multicolumn{1}{|c|}{ }  & \multicolumn{4}{c|}{Inferred $R_c$ (realistic errors)} \\
\hline
\multirow{2}{*}{Model} & \multicolumn{2}{c||}{$R_c=R_H$} & \multicolumn{2}{c|}{$R_c=4R_H$} \\
\cline{2-5}
 & $n=1$ & $n=10$ & $n=1$ & $n=10$  \\
\hline
 \multirow{3}{*}{$D=5$}&  $1.25^{+10.25}_{-0.84}$  & $1.04^{+0.54}_{-0.30}$&$5.53^{+8.98}_{-4.86}$ & $4.90^{+7.58}_{-2.41}$ \\
 \cdashline{2-5}
  & $0.95^{+5.65}_{-0.50}$  & $1.13^{+1.03}_{-0.33}$&$5.53^{+8.94}_{-4.71}$ & $4.92^{+8.28}_{-1.98}$ \\
\cdashline{2-5}
& $1.04^{+4.55}_{-0.69}$  & $1.01^{+0.32}_{-0.22}$&$6.17^{+8.37}_{-5.08}$ & $4.62^{+7.93}_{- 1.35}$   \\
\hline
\multirow{3}{*}{$D=6$}&  $0.98^{+2.75}_{-0.56}$  & $1.02^{+0.24}_{-0.17}$&$6.26^{+8.27}_{-5.18}$ & $4.64^{+5.43}_{-1.60}$ \\
\cdashline{2-5}
 & $0.96^{+2.69}_{-0.55}$  & $1.00^{+0.27}_{-0.17}$&$5.50^{+8.88}_{-4.50}$ & $4.09^{+1.61}_{-0.79}$  \\
\cdashline{2-5}
 & $1.08^{+1.73}_{-0.53}$  & $1.00^{+0.16}_{-0.12}$&$6.06^{+8.37}_{-4.44}$ & $4.00^{+1.13}_{-0.68}$ \\
\hline 
\multirow{3}{*}{$D=7$}&  $1.04^{+1.86}_{-0.53}$  & $1.03^{+0.17}_{-0.12}$&$5.80^{+8.50}_{-4.11}$ & $4.28^{+1.97}_{-0.76}$  \\
\cdashline{2-5}
 & $1.05^{+1.63}_{-0.48}$  & $1.01^{+0.16}_{-0.11}$&$4.68^{+8.74}_{-3.27}$ & $4.11^{+0.93}_{-0.58}$ \\
\cdashline{2-5}
& $1.04^{+1.14}_{-0.42}$  & $1.00^{+0.11}_{-0.09}$&$6.21^{+7.69}_{-3.99}$ & $4.07^{+0.64}_{-0.45}$\\
\hline
\end{tabular}
\end{center}
\vspace{8mm}

\end{table}

\subsection{Constraints on the screening scale} 
We now consider LISA's ability to place limits on the screening scale. The median FoMs and $95\%$ CIs for all MBHB formation models can be found in Table \ref{PEstatsR}. This Table and Figure \ref{fig:contDn} show that, just like for the constraints on the number of dimensions, the constraints on the screening scale improve the steeper the transition and the lower the screening scale. However, unlike for the number of dimensions, it appears that LISA's capability to constrain the screening scale strongly depends on the hypothetical true value of $D$ and which galaxy formation model is considered. Furthermore, in the case of realistic errors the constraints on $R_c$ for {\em all\/} models are substantially weaker.

With optimistic siren errors and for the $D=5$ cosmological model we see that LISA is only capable of constraining the screening scale in the most favourable scenario $(n=10, R_c=R_H)$ of the Q3nod and Q3d model, and even then with a precision of $14.3\%$ and $19.6\%$ respectively. As the number of dimensions increases, however, the FoMs tend towards their true values and their error becomes significantly smaller for all the MBHB formation models. Moreover, for the $D=6$ model and with optimistic siren errors, one can place meaningful constraints on the screening scale in all cosmological scenarios and MBHB formation models, with a precision of $8.8\%$ for the most favourable scenario and Q3nod model -- except for the $(n=1, R_c=4R_H)$ as one would expect since it is the most challenging scenario to constrain.

Finally, for $D=7$, we see that even with optimistic siren errors the $n=1, R_c=4R_H$ model parameters are still poorly constrained but the constraints on the screening scale are much better for the other cosmological models considered and for all three galaxy formation models, particularly when $n=10$. For $n=10, R_c=R_H$ and the Q3nod model, for example, the error on $R_c$ is only $7.2\%$. Again these results clearly show the strong improvement that comes with a larger sample of sirens. In particular, the constraints on the screening scale are much tighter for the Q3nod model, where the number of standard sirens is considerably higher, than in the other two galaxy formation models. 


\subsection{Constraints on the steepness parameter and Hubble parameter} 
Tables \ref{PEn}-\ref{PEH0} present the median FoMs and CIs for the cosmological parameters $n$ and $H_0$  for all MBHB formation scenarios and for all of the cosmological models considered. These results will not be discussed further since they are fully consistent with the results obtained for $D$ and $R_c$. Note however that the results from Table \ref{PEH0} imply that LISA could in principle provide a fully independent constraint on the Hubble constant, complementing the other optical measurements. Although the errors on $H_0$ are slightly higher compared to other cosmological probes, even in the most optimistic cosmological and MBHB formation scenario, it is still important to note that LISA could potentially constrain this parameter, which cannot be independently measured using {\em only\/} observations of the Pantheon supernova sample.

\begin{table}[ht]
\begin{center}
\small
\setlength{\tabcolsep}{4.0pt} 
\setlength\abovecaptionskip{-5pt}
 \setlength\belowcaptionskip{0pt}
\caption{\label{PEn} 
Median figures of merit and $95\%$ credible intervals summarising the marginalised posterior PDF of the steepness parameter, $n$,for the various assumed true cosmological scenarios and MBHB formation models considered, for both optimistic (top panel) and realistic (bottom panel) errors.}
\vspace{4mm}

\begin{tabular}{|c|c|c||c|c|}
\hline 
\multicolumn{1}{|c|}{ } & \multicolumn{4}{c|}{Inferred $n$ (optimistic errors)} \\
\hline
\multirow{2}{*}{Model} & \multicolumn{2}{c||}{$R_c=R_H$} & 
    \multicolumn{2}{c|}{$R_c=4R_H$}  \\
\cline{2-5}
 & $n=1$ & $n=10$ & $n=1$ & $n=10$  \\
\hline
 \multirow{3}{*}{$D=5$}& $1.02^{+1.79}_{-0.31}$  & $9.57^{+5.09}_{-6.05}$&$0.96^{+0.85}_{-0.26}$ & $7.83^{+6.67}_{-5.71}$   \\
 \cdashline{2-5}
 & $0.99^{+1.12}_{-0.31}$  & $9.24^{+5.42}_{-6.35}$&$0.98^{+0.64}_{-0.21}$ &$8.62^{+5.96}_{-5.95}$  \\
\cdashline{2-5}
& $0.99^{+0.63}_{-0.24}$  & $9.87^{+4.82}_{-5.76}$&$0.96^{+0.45}_{-0.20}$ &$8.98^{+5.67}_{-5.54}$  \\
\hline
\multirow{3}{*}{$D=6$}& $1.01^{+0.55}_{-0.22}$  & $9.95^{+4.76}_{-5.48}$&$0.94^{+0.35}_{-0.17}$ &$9.66^{+4.97}_{-5.43}$  \\
\cdashline{2-5}
 & $1.03^{+0.65}_{-0.25}$  & $9.35^{+5.30}_{-6.15}$&$0.97^{+0.39}_{-0.18}$ &$9.24^{+5.38}_{-5.10}$   \\
\cdashline{2-5}
&  $0.95^{+0.28}_{-0.16}$  & $9.85^{+4.70}_{-4.76}$&$0.97^{+0.27}_{-0.15}$ &$10.08^{+4.64}_{-4.86}$ \\
\hline 
\multirow{3}{*}{$D=7$}& $0.98^{+0.32}_{-0.17}$  & $9.90^{+4.78}_{-4.87}$&$0.96^{+0.25}_{-0.14}$ &$10.27^{+4.48}_{-4.86}$   \\
\cdashline{2-5}
 & $1.01^{+0.42}_{-0.20}$  & $9.48^{+4.98}_{-5.45}$&$0.97^{+0.23}_{-0.14}$ &$10.33^{+4.40}_{-4.43}$   \\
\cdashline{2-5}
& $1.01^{+0.19}_{-0.14}$  & $10.50^{+4.17}_{-4.15}$&$0.96^{+0.18}_{-0.12}$ &$9.94^{+4.71}_{-3.88}$ \\
\hline
\end{tabular}

\begin{tabular}{|c|c|c||c|c|}
\hline 
\multicolumn{1}{|c|}{ } & \multicolumn{4}{c|}{Inferred $n$ (realistic errors)} \\
\hline
\multirow{2}{*}{Model} & \multicolumn{2}{c||}{$R_c=R_H$} & 
    \multicolumn{2}{c|}{$R_c=4R_H$}  \\
\cline{2-5}
 & $n=1$ & $n=10$ & $n=1$ & $n=10$  \\
\hline
 \multirow{3}{*}{$D=5$}& $0.94^{+8.44}_{-0.38}$  & $8.94^{+5.70}_{-6.87}$&$1.05^{+3.90}_{-0.39}$ & $7.06^{+7.48}_{-5.23}$   \\
 \cdashline{2-5}
 & $1.43^{+7.11}_{-0.73}$  & $8.03^{+6.56}_{-6.50}$&$0.93^{+0.81}_{-0.25}$ &$7.01^{+7.51}_{-5.17}$  \\
\cdashline{2-5}
& $0.98^{+5.86}_{-0.34}$  & $9.12^{+5.50}_{-6.55}$&$1.02^{+4.35}_{-0.34}$ &$7.32^{+7.19}_{-5.38}$  \\
\hline
\multirow{3}{*}{$D=6$}& $1.01^{+7.88}_{-0.32}$  & $9.41^{+5.22}_{-5.84}$&$0.92^{+0.72}_{-0.21}$ &$7.23^{+7.29}_{-5.07}$  \\
\cdashline{2-5}
 & $0.99^{+4.55}_{-0.33}$  & $8.78^{+5.85}_{-6.42}$&$0.94^{+0.80}_{-0.21}$ &$9.98^{+4.76}_{-6.03}$   \\
\cdashline{2-5}
&  $0.97^{+0.75}_{-0.26}$  & $9.51^{+5.14}_{-5.32}$&$0.94^{+0.45}_{-0.17}$ &$9.85^{+4.86}_{-5.94}$ \\
\hline 
\multirow{3}{*}{$D=7$}& $1.00^{+0.74}_{-0.23}$  & $9.56^{+5.09}_{-5.20}$&$0.95^{+0.50}_{-0.17}$ &$8.74^{+5.87}_{-5.54}$   \\
\cdashline{2-5}
 & $1.00^{+2.01}_{-0.28}$  & $9.43^{+5.21}_{-6.41}$&$0.98^{+0.65}_{-0.20}$ &$9.70^{+4.97}_{-5.56}$   \\
\cdashline{2-5}
& $0.99^{+0.46}_{-0.22}$  & $9.70^{+4.98}_{-5.23}$&$0.93^{+0.32}_{-0.14}$ &$10.27^{+4.50}_{-5.26}$ \\
\hline
\end{tabular}
\end{center}
\end{table}

\begin{table}[ht] 
\begin{center}
\small
\setlength{\tabcolsep}{4.0pt} 
\setlength\abovecaptionskip{-5pt}
 \setlength\belowcaptionskip{0pt}
\caption{\label{PEH0} 
Median figures of merit and $95\%$ credible intervals summarising the marginalised posterior PDF of the Hubble constant, $H_0$, for the various assumed true cosmological scenarios and MBHB formation models considered, for both optimistic (top panel) and realistic (bottom panel) errors.}
\vspace{4mm}

\footnotesize
\begin{tabular}{|c|c|c||c|c|}
\hline 
\multicolumn{1}{|c|}{ } &\multicolumn{4}{c|}{Inferred $H_0$ (optimistic errors)} \\
\hline
\multirow{2}{*}{Model} & \multicolumn{2}{c||}{$R_c=R_H$} & 
    \multicolumn{2}{c|}{$R_c=4R_H$}   \\
\cline{2-5}
 & $n=1$ & $n=10$ & $n=1$ & $n=10$  \\
\hline
 \multirow{3}{*}{$D=5$}& $73.16^{+3.23}_{-3.49}$& $73.23^{+3.24}_{-3.14}$  & $73.14^{+3.23}_{-3.38}$&$73.43^{+2.24}_{-1.89}$  \\
 \cdashline{2-5}
 & $73.19^{+3.27}_{-3.50}$  &  $73.22^{+3.26}_{-3.73}$&$73.10^{+3.22}_{-3.42}$ & $73.26^{+2.31}_{-1.91}$ \\
\cdashline{2-5}
& $73.21^{+3.23}_{-3.47}$ & $73.24^{+2.96}_{-2.69}$&$73.14^{+3.17}_{-3.21}$ & $73.34^{+1.69}_{-1.36}$   \\
\hline
\multirow{3}{*}{$D=6$}&  $73.23^{+3.29}_{-3.39}$  & $73.24^{+3.24}_{-3.12}$&$73.16^{+3.22}_{-3.35}$ & $73.40^{+2.12}_{-1.83}$  \\
\cdashline{2-5}
 & $73.22^{+3.27}_{-3.42}$  & $73.23^{+3.25}_{-3.38}$&$73.22^{+3.31}_{-3.37}$ & $73.29^{+2.09}_{-1.86}$  \\
\cdashline{2-5}
 & $73.22^{+3.26}_{-3.41}$  & $73.24^{+3.07}_{-2.80}$&$73.21^{+3.26}_{-3.17}$ & $73.59^{+1.50}_{-1.41}$\\
\hline 
\multirow{3}{*}{$D=7$}&  $73.23^{+3.28}_{-3.38}$  & $73.24^{+3.24}_{-3.30}$&$73.20^{+3.27}_{-3.36}$ & $73.23^{+1.89}_{-1.82}$   \\
\cdashline{2-5}
 &  $73.24^{+3.29}_{-3.37}$  & $73.23^{+3.28}_{-3.39}$&$73.24^{+3.22}_{-3.35}$ & $73.28^{+1.97}_{-1.75}$  \\
\cdashline{2-5}
&  $73.24^{+3.28}_{-3.34}$  & $73.24^{+3.02}_{-3.02}$&$73.23^{+3.28}_{-3.29}$ & $73.19^{+1.42}_{-1.34}$\\
\hline
\end{tabular}

\begin{tabular}{|c|c|c||c|c|}
\hline 
\multicolumn{1}{|c|}{ } &\multicolumn{4}{c|}{Inferred $H_0$ (realistic errors)} \\
\hline
\multirow{2}{*}{Model} & \multicolumn{2}{c||}{$R_c=R_H$} & 
    \multicolumn{2}{c|}{$R_c=4R_H$}   \\
\cline{2-5}
 & $n=1$ & $n=10$ & $n=1$ & $n=10$  \\
\hline
 \multirow{3}{*}{$D=5$}& $73.18^{+3.29}_{-3.45}$& $73.22^{+3.28}_{-3.34}$  & $73.07^{+3.25}_{-3.47}$&$73.24^{+2.99}_{-2.79}$  \\
 \cdashline{2-5}
 & $73.20^{+3.29}_{-3.46}$  &  $73.18^{+3.31}_{-3.48}$&$73.12^{+3.30}_{-3.47}$ & $73.10^{+3.11}_{-2.90}$ \\
\cdashline{2-5}
& $73.19^{+3.27}_{-3.47}$ & $73.22^{+3.24}_{-3.20}$&$73.09^{+3.26}_{-3.50}$ & $73.28^{+2.69}_{-2.47}$   \\
\hline
\multirow{3}{*}{$D=6$}&  $73.22^{+3.32}_{-3.44}$  & $73.23^{+3.27}_{-3.32}$&$73.20^{+3.28}_{-3.41}$ & $73.19^{+2.95}_{-2.81}$  \\
\cdashline{2-5}
 & $73.23^{+3.27}_{-3.42}$  & $73.22^{+3.27}_{-3.42}$&$73.21^{+3.30}_{-3.41}$ & $73.29^{+2.99}_{-2.81}$  \\
\cdashline{2-5}
 & $73.22^{+3.27}_{-3.43}$  & $73.23^{+3.22}_{-3.24}$&$73.22^{+3.27}_{-3.37}$ & $73.34^{+2.61}_{-2.49}$\\
\hline 
\multirow{3}{*}{$D=7$}&  $73.23^{+3.30}_{-3.41}$  & $73.23^{+3.29}_{-3.35}$&$73.24^{+3.31}_{-3.40}$ & $73.21^{+2.98}_{-2.81}$   \\
\cdashline{2-5}
 &  $73.23^{+3.29}_{-3.40}$  & $73.24^{+3.30}_{-3.38}$&$73.24^{+3.34}_{-3.42}$ & $73.30^{+3.10}_{-2.91}$  \\
\cdashline{2-5}
&  $73.23^{+3.29}_{-3.39}$  & $73.24^{+3.27}_{-3.28}$&$73.12^{+3.25}_{-3.39}$ & $73.14^{+2.50}_{-2.51}$\\
\hline
\end{tabular}
\end{center}
\vspace{8mm}

\end{table}
\normalsize

\newpage 
\begin{figure}[ht] 
\includegraphics[width=1.\textwidth]{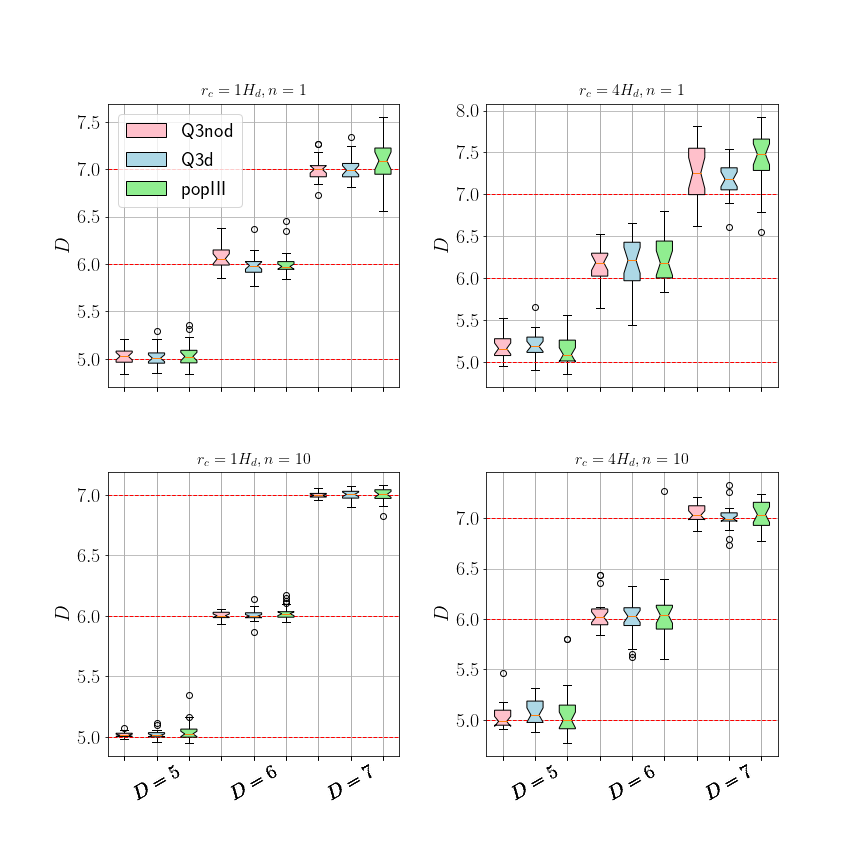}
\caption{\label{fig:boxD_opt} Boxplot distributions for the median value of the marginalised posterior PDF of the number of dimensions, $D$, assuming optimistic errors. Results are shown for all 22 catalogues analysed in each of the three MBHB formation scenarios, and for a number of different true cosmological models.}
\end{figure}

\begin{figure}[ht]
\includegraphics[width=1.\textwidth]{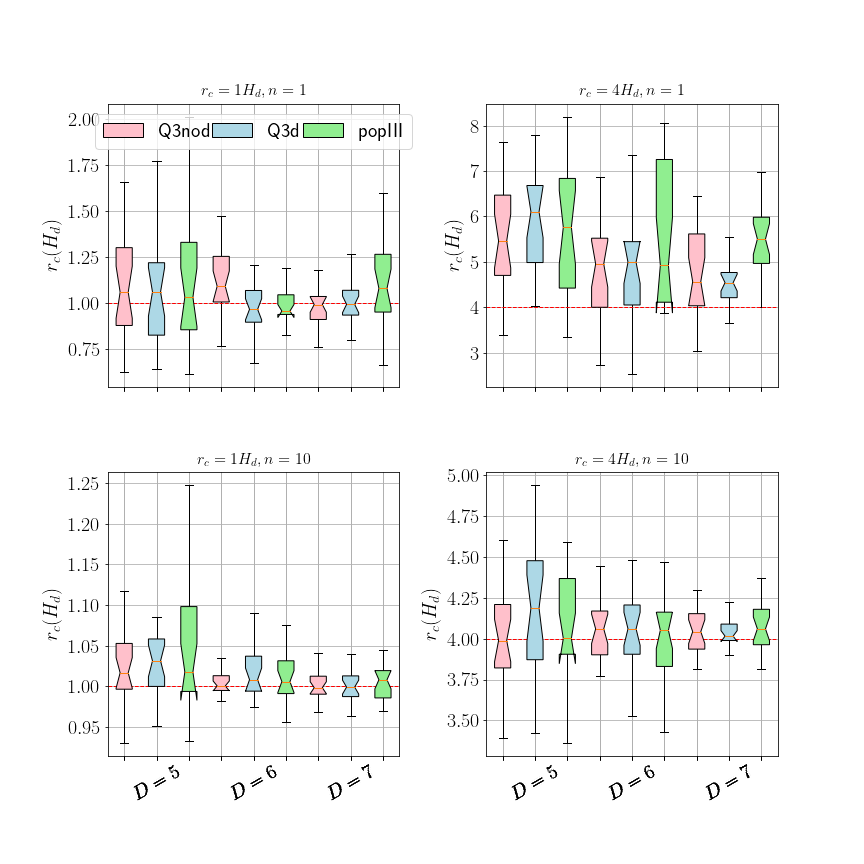}
\caption{\label{fig:boxR_opt} 
Boxplot distributions for the median value of the marginalised posterior PDF of the screening scale parameter, $r_c$, assuming optimistic errors. Results are shown for all 22 catalogues analysed in each of the three MBHB formation scenarios, and for a number of different true cosmological models. Note that the outliers were omitted for clarity, but the number of these decreases the steeper the transition and the lower the screening scale.}
\end{figure}

\subsection{Statistical spread of the catalogues} 
Although we employed 22 synthetic catalogues for each MBHB formation model in order to improve the statistics of our analysis, the scatter of the properties of the massive binary black hole population across the catalogues is still quite high. This can be seen by considering the distribution of the median values of the marginalised PDFs for the inferred cosmological parameters. In Figures \ref{fig:boxD_opt} and \ref{fig:boxR_opt} we present example box plots showing the distribution of median values for the inferred cosmological dimension $D$ and screening scale $R_C$ respectively, from all 22 catalogues and for all three MBHB formation scenarios, assuming several different true cosmological models. As one would expect from our results in Section \ref{resultsMCMC}, we see from Figure \ref{fig:boxD_opt} that the median value of $D$ is closer to the assumed true value, and with smaller spread, for the models with higher steepness parameter and lower screening scale. Also, the distribution of the median values of the screening scale is more peaked as the number of dimensions increases, although the spread in the median values of $D$ is not very sensitive to the number of dimensions, in agreement with our earlier results. Finally, for any cosmological model the Q3nod model gives systematically better results than the other two models -- as a result of more BH binaries being formed in this model.
 

\subsection{Bayesian model selection results}

To quantify LISA's ability to constrain higher-dimensional theories with a given screening scale and steepness parameter we calculate, for each non-GR model and for every catalogue and MBHB formation model, the evidence, $Z$, and hence the log Bayes factor, $B_{\rm NGR/DGP}={Z_{\rm NGR}}/{Z_{\rm DGP}}$, as given in eq. \ref{bayesfact}. We then choose the median of these computed values to represent the log Bayes factor between the NGR and DGP models. Moreover, to estimate the error in the log Bayes factor we use the $95\%$ credible interval around the median value for all of the catalogue realizations. The reason for this is that the formal error on the log Bayes factor for a {\em single\/} catalogue realisation is considerably smaller than the scatter in the values across the 22 generated catalogues within each MBHB formation model.
Table \ref{Bfactor} shows the results of the model comparison of all the alternative non GR models considered against the DGP model, for all MBHB formation scenarios, and assuming either `optimistic' or `realistic' errors. \\

\begin{table}[ht]
\begin{center}
\small
\setlength{\tabcolsep}{4.0pt} 
\setlength\abovecaptionskip{-5pt}
 \setlength\belowcaptionskip{0pt}
\caption{\label{Bfactor} Median figures of merit and $95\%$ credible intervals summarising the log Bayes factor for a particular NGR model, compared with the DGP model, for the various cosmological scenarios and MBHB formation models considered. In each row of the table separated by dashed lines, the top sub-row shows the FoMs and CI for light-seeds (popIII), the central sub-row for heavy seeds with delays (Q3d) and the bottom sub-row for heavy seeds without delays (Q3nod). The Hubble radius $R_H=4.093$ Gpc. A positive log Bayes factor implies evidence in favour of the NGR model.}
\vspace{3mm}

\begin{tabular}{|c|c|c||c|c|}
\hline 
\multicolumn{1}{|c|}{ } & \multicolumn{4}{c|}
{$\log B_{\rm NGR/DGP}$ (optimistic errors)} \\
\hline
\multirow{2}{*}{Model} & \multicolumn{2}{c||}{$R_c=R_H$} & 
    \multicolumn{2}{c|}{$R_c=4R_H$}  \\
\cline{2-5}
 & $n=1$ & $n=10$ & $n=1$ & $n=10$ \\
\hline
 \multirow{3}{*}{$D=6$}&  $5.3^{+9.20}_{-13.3}$& $15.3^{+12.0}_{-11.5}$  & $2.08^{+12.6}_{-7.92}$ &$3.52^{+17.5}_{-14.2}$  \\
 \cdashline{2-5}
 &$3.08^{+13.52}_{-11.2}$& $4.56^{+10.0}_{-16.1}$  &$1.67^{+12.2}_{-12.9}$&  $3.08^{+10.9}_{-13.1}$   \\
\cdashline{2-5}
&$20.0^{+9.64}_{-12.78}$& $24.2^{+16.1}_{-7.26}$  & $9.01^{+9.39}_{-12.04}$&  $21.7^{+12.73}_{-15.2}$    \\
\hline
\multirow{3}{*}{$D=7$}& $10.6^{+6.57}_{-9.12}$ & $25.5^{+15.1}_{-7.49}$  & $9.40^{+9.16}_{-9.18}$  & $13.6^{+19.1}_{-9.03}$  \\
\cdashline{2-5}
 &$12.6^{+11.62}_{-13.9}$& $24.1^{+13.4}_{-7.80}$  & $6.42^{+13.2}_{-13.7}$&  $11.1^{+8.4}_{-7.13}$   \\
\cdashline{2-5}
 & $15.3^{+14.6}_{-19.65}$&$26.0^{+14.5}_{-22.3}$&  $9.90^{+12.9}_{-22.02}$ &$11.2^{+9.74}_{-17.7}$  \\
\hline
\end{tabular}
 \vspace{+1.0em}

\begin{tabular}{|c|c|c||c|c|}
\hline 
\multicolumn{1}{|c|}{ } & \multicolumn{4}{c|}
{$\log B_{\rm NGR/DGP}$ (realistic errors)} \\
\hline
\multirow{2}{*}{Model} & \multicolumn{2}{c||}{$R_c=R_H$} & 
    \multicolumn{2}{c|}{$R_c=4R_H$}  \\
\cline{2-5}
 & $n=1$ & $n=10$ & $n=1$ & $n=10$ \\
\hline
 \multirow{3}{*}{$D=6$}&  $2.07^{+12.2}_{-7.87}$& $3.23^{+8.1}_{-7.97}$  & $0.20^{+13.6}_{-8.25}$ &$3.10^{+8.09}_{-7.56}$  \\
 \cdashline{2-5}
 &$1.81^{+20.3}_{-7.85}$& $1.87^{+12.3}_{-14.4}$  &$-0.68^{+12.8}_{-3.89}$&  $3.07^{+11.68}_{-10.13}$   \\
\cdashline{2-5}
&$5.84^{+7.21}_{-9.82}$& $7.16^{+17.1}_{-26.4}$  & $1.3^{+15.2}_{-13.7}$&  $5.2^{+15.8}_{-23.9}$    \\
\hline
\multirow{3}{*}{$D=7$}& $5.20^{+15.3}_{-9.67}$ & $5.80^{+7.03}_{-12.04}$  & $0.34^{+16.2}_{-13.0}$  & $3.07^{+10.13}_{-5.67}$  \\
\cdashline{2-5}
 &$6.10^{+10.7}_{-9.73}$& $5.72^{+5.51}_{-10.11}$  & $-1.07^{+13.9}_{-12.4}$&  $-1.01^{+7.43}_{-10.66}$   \\
\cdashline{2-5}
 & $5.60^{+14.81}_{-7.48}$&$12.8^{+15.8}_{-17.9}$& $1.90^{+13.2}_{-17.9}$ &$3.35^{+8.65}_{-9.73}$  \\
\hline
\end{tabular}
\end{center}
\end{table}

A log Bayes factor that is greater than $\sim 5$ indicates strong evidence in favour of the NGR model. Since our data were simulated for NGR cosmological models, where gravity is modified according to eq. (\ref{hNGR}), we would expect the data to favour the NGR model over the DGP model in the cases considered here.  

The first thing to note from Table \ref{Bfactor} is that, as expected, the Q3nod model gives systematically higher evidence in favor of the true NGR model than do the other two MBHB formation models -- again confirming the importance of the number of sources that LISA will observe in determining its discriminatory power. Nevertheless we see from the left column of Table \ref{Bfactor} that, for the case of optimistic errors, when $D=7$ the NGR models are indeed favored for all three MBHB formation models and for each cosmological scenario considered -- although, as expected, the evidence is stronger when $n=10$. When $D=6$ the NGR model remains strongly favored in the Q3nod formation scenario, and also for the popIII case when $R_c=R_H$. When $R_c=4 R_H$, however, the evidence favoring NGR is weaker -- particularly for the Q3d formation scenario.

The pattern of our results is broadly similar for the case of realistic errors, as can be seen in the right column of Table \ref{Bfactor} -- although the log Bayes factor values are substantially smaller in all cases.  Nevertheless, the NGR model is still generally favored (albeit with a larger spread of log Bayes factors across the 22 catalogues) for all MBHB formation scenarios provided $R_c= R_H$ and $D=7$. When $R_c= 4 R_H$, however, the larger scatter of the realistic errors means that the evidence for the NGR model is no longer conclusive.

\subsection{Constraints on the DGP model using the full LISA and Pantheon samples}

Finally we can perform a statistical analysis for the Pantheon and LISA data combined, using the full seven-dimensional parameter space of the DGP model (i.e. the 4 DGP parameters discussed in Section \ref{section2}, augmented by the parameters $D$, $R_c$ and $n$ associated with the \emph{ansatz} of eq. \ref{hNGR}) for the catalogues corresponding to the different MBHB formation scenarios. We illustrate this joint analysis for two cases, each for true parameters $D=5$ and $R_c=1.2R_H$, and assuming true values of $n=1$ and $n=10$ respectively. The credible regions on the inferred Hubble parameter from this joint analysis are summarised in Table \ref{tab:PEstatsR}, and in the case of optimistic errors (left panel) and realistic errors (right panel) respectively. (Note that, as in Section \ref{sec:lumdis}, a value of ${\cal M} = -19.3$ was adopted for the Pantheon sample).  We see that the median values and credible regions for $H_0$ are generally very similar to those presented in Table \ref{tab:Pantheon}, where only the Pantheon data were fitted to the DGP model.

\begin{table}[ht]
\begin{center}
\setlength{\tabcolsep}{4.0pt} 
 \setlength\belowcaptionskip{0pt}
\caption{\label{tab:PEstatsR} Median figures of merit and $95\%$ credible intervals summarising the marginalised posterior PDFs of the Hubble parameter for the various MBHB formation models, when considering constraints on the full seven-dimensional DGP model using a combined data set of Pantheon and LISA data. 
}

\begin{tabular}{|c|c||c|}
\hline 
\multicolumn{1}{|c|}{Optimistic errors}  & \multicolumn{2}{c|}{$H_0(\text{kms}^{-1} \text{Mpc}^{-1}$)} \\
\hline
 \multicolumn{1}{|c|}{Model}& $n=1$ & $n=10$   \\
\hline
 \multirow{3}{*}{$D=5$}&  $71.57^{+0.89}_{-0.88}$  & $71.49^{+0.78}_{-0.73}$ \\
 \cdashline{2-3}
  & $71.55^{+0.87}_{-0.87}$  & $71.42^{+0.75}_{-0.71}$\\
\cdashline{2-3}
& $71.61^{+0.85}_{-0.86}$  & $71.38^{+0.71}_{-0.67}$ \\
\hline
\end{tabular}

\begin{tabular}{|c|c||c|}
\hline 
\multicolumn{1}{|c|}{Realistic errors}  & \multicolumn{2}{c|}{$H_0(\text{kms}^{-1} \text{Mpc}^{-1}$)} \\
\hline
 \multicolumn{1}{|c|}{Model}& $n=1$ & $n=10$   \\
\hline
 \multirow{3}{*}{$D=5$}&  $71.60^{+0.88}_{-0.90}$  & $71.49^{+0.86}_{-0.83}$ \\
 \cdashline{2-3}
  & $71.59^{+0.86}_{-0.86}$  & $71.47^{+0.86}_{-0.83}$\\
\cdashline{2-3}
& $71.56^{+0.86}_{-0.84}$  & $71.47^{+0.81}_{-0.77}$ \\
\hline
\end{tabular}
\end{center}
\end{table}

\begin{figure}[ht]
\includegraphics[width=1\textwidth]{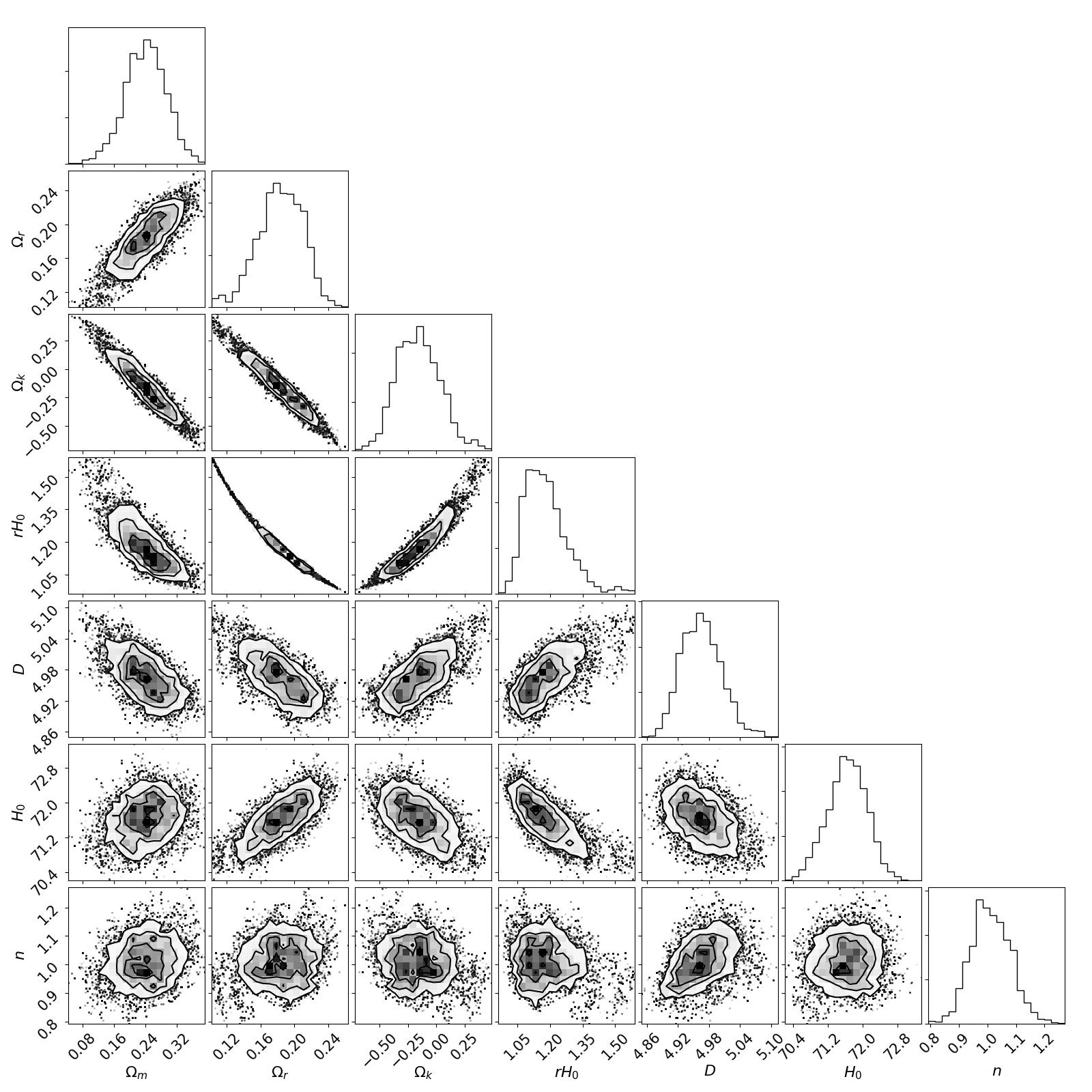}
\caption{\label{fig:DGPLISA} Joint posterior PDF for the full seven-dimensional parameter space -- comprising the four DGP parameters augmented by $D$, $R_c$ and $n$ -- analysing Pantheon and LISA data combined. The analysis is for a representative catalogue in the Q3d formation scenario and assuming optimistic errors. True parameter values $D=5$,$R_c=1.2R_H$,$n=1$ were adopted and a value of ${\cal M} = -19.3$ was adopted for the Pantheon sample.}
\end{figure}

The joint posterior distribution inferred for full seven-dimensional DGP model parameter space is also shown in Figure \ref{fig:DGPLISA}, for a representative Q3d formation catalogue and assuming optimistic errors.  Again, we note from this Figure that the constraints on the Hubble parameter are comparable to those shown in Figure \ref{fig:pantheon_test}, where only constraints on the four DGP parameters, using only the Pantheon data, were considered.  Thus we see that in principle LISA will be capable of placing tight constraints on the Hubble parameter while {\em simultaneously\/} constraining the parameters of non-GR models such as those considered in our analysis. 

\section{Conclusions} 
\label{conclusion}
In this paper we have investigated the potential capability of the LISA satellite to place constraints on higher-dimensional cosmological theories with non-compact spacetime dimensions, by using GW standard sirens observed with an EM counterpart. In the absence of a complete, unique GW model for these theories, we used a phenomenological {\em ansatz\/} for the GW amplitude, which is based on the physics of the DGP model with positive branch solutions and includes a screening scale $R_c$ beyond which the GWs leak into the higher dimensions, leading to a reduction in the amplitude of the observed GWs and hence a systematic error in the inferred distance to the source. 

Considering various plausible cosmological scenarios (number of dimensions $D= 5, 6$ or $7$; steepness parameter $n=1$ or $10$ and screening scale $R_c=R_H \sim 4$Gpc or $4R_H$) and three models for the MBHB formation (``heavy seeds without delays"; ``heavy seeds with delays" and ``popIII stars") we have investigated catalogues of MBHB events for which an optical counterpart may plausibly be expected, so that each event may be assigned a GW luminosity distance, the redshift of corresponding EM counterpart and an estimate of the error on measured distance. These catalogues are based on those previously presented in \cite{eLISA,eLISA2} where they were used to investigate LISA constraints on other cosmological models.

We have found that, in general, the heavy seeds with no delays model (where the number of detectable standard sirens amounts to a four-year average of $\sim 27$) gives systematically better model constraints than the other two scenarios, where the four-year average only amounts to about $12$ and $14$ sirens for the popIII and Q3nod respectively. Thus, we conclude that the ability of LISA to place meaningful constraints on the number of spacetime dimensions and screening scale will strongly depend on the actual number and redshift distribution of MBHB merger events, and the corresponding efficiency in identifying a host galaxy redshift. Furthermore, by considering two different scenarios, denoted `optimistic' and `realistic', for the total error on the luminosity distance -- derived from a combination of the GW observations, the impact of peculiar velocities and weak lensing, and the uncertainty on the redshift of the counterpart -- we concluded that the constraints on the parameters will strongly depend on the error with which we can measure the redshift of the sources, the constraints being substantially weaker for the realistic error scenario.

We also found that LISA's ability to constrain higher-dimensional theories, as defined by the scaling relation in (\ref{hNGR}), will depend on the cosmological parameters defining the theory -- namely the number of dimensions, screening scale and transition steepness -- with the $(D=7, R_c=R_H, n=10)$ scenario, where modification from GR is the most pronounced, being much better constrained than the other cases.

In the model where the modification to gravity is the least pronounced, namely $(R_c= 4R_H, n=1)$, we found that LISA is not sensitive enough to provide meaningful constraints on the parameters. This is the case for $D=5, 6$ and $7$ and all MBHB formation models, although the popIII and Q3d scenarios give slightly worse results. On the other hand, for any of the other cosmological models and error scenarios considered, we found that LISA will be able to place meaningful constraints on the number of spacetime dimensions. Moreover, this constraining power considerably improves not only for greater numbers of standard sirens observed but also the steeper the transition and the smaller the screening scale -- reaching a precision at the level of $0.86 \%$ for the most favorable cosmological scenario and formation model.

We found that the screening scale $R_c$ is not as well constrained as $D$, but that unlike the constraints on $D$, the constraints on $R_c$ are very sensitive to the true number of spacetime dimensions. Indeed, LISA is unable to place meaningful constraints on $R_c$ when $D=5$, unless $(n=10, R_c=R_H)$ and the Q3nod or Q3 model is considered, but for $D \geq 6 $, except for the least favourable cosmological scenario, LISA can place meaningful constraints on $R_c$ in all cosmological models and MBHB formation scenarios reaching a precision of $7.2\%$ for $(D=7,n=10, R_c=R_H)$ within the Q3nod model for the realistic error scenario.

Finally, we investigated how strongly the siren data would favour the true model of modified gravity from which it was simulated, when compared with the DGP model. To that end we calculated the Bayesian evidence in favour of each higher-dimensional model considered, and found that for the case of optimistic siren errors the log evidence favored the true NGR model in every case. For realistic siren errors the evidence generally also favored the true NGR model -- although the log Bayes factor values were substantially smaller and with larger scatter.

In summary, we have shown that standard sirens observed with LISA in the redshift range $1<z<8$ have the potential to test higher-dimensional theories with non compact extra-dimensions in a completely different way than current EM probes. However, our analysis is a phenomenological one, modelling the GW damping by considering a very general type of leakage for large extra dimensions. Our results do not hold for higher-dimensional theories with compact extra dimensions such as string theory. Additionally, even though the GW amplitude scaling used applies to the DGP model, the so-called `infrared transparency' effect can be shown to result in a distance at which GW damping is manifested in the DGP model much beyond the distances to sources observable with frequencies relevant to LISA \cite{IR}. Nevertheless, our analysis still provides a useful measure of the constraints we can place on extra-dimensional theories with large $(\geq 100  \text{km})$ extra dimensions. 


\acknowledgments
CE-R is supported by the Royal Astronomical Society as FRAS 10147, PAPIIT-UNAM Project IA100220.
M. H. is supported by the Science and Technology Facilities Council (Ref. ST/L000946/1). M.C. is supported by the Perimeter Institute for Theoretical Physics.  The authors gratefully acknowledge the help of Alberto Sesana and Nicola Tamanini, and their co-authors, for providing access to the mock LISA siren catalogues used in this paper.

%

\bibliographystyle{unsrt}
\bibliography{LISA}
\end{document}